\documentclass[fleqn,11pt]{wlscirep}
\usepackage{setspace}
\usepackage[utf8]{inputenc}
\usepackage[T1]{fontenc}
\usepackage{bm}
\usepackage{csquotes}
\usepackage{makecell}
\usepackage{sectsty}
\usepackage{amssymb}
\usepackage{amsmath}
\usepackage{mathtools}
\usepackage{booktabs}
\sectionfont{\large}
\subsectionfont{\normalsize}
\usepackage{float}
\usepackage{array}
\usepackage{colortbl}
\usepackage{tablefootnote}
\definecolor{lightgray}{RGB}{230, 230, 230}
\usepackage[mathscr]{euscript}
\usepackage{hyperref}
\hypersetup{linkcolor=black}

\usepackage[style=ieee, citestyle=numeric-comp, sorting = none]{biblatex}
\usepackage[format=plain, justification=justified,font={sf}]{caption}
\usepackage{newfloat}
\usepackage{wrapfig}

\usepackage{mathtools}

\usepackage{gensymb}
\usepackage{mathtools, nccmath}

\usepackage{titlesec}

\usepackage{cjhebrew}

\usepackage{lineno}
\usepackage{fancyhdr}

\sectionfont{\fontsize{15}{18}\selectfont}
\subsectionfont{\fontsize{13}{16}\selectfont}
\subsubsectionfont{\fontsize{12}{15}\selectfont}

\DeclareUnicodeCharacter{0301}{\'{e}}

\begin{document}

\begin{titlepage}
    \begin{center}
        \textbf{\fontsize{15}{18}\selectfont Models of Animal Behavior \\\vspace{3pt}as Active Particle Systems with Nonreciprocal Interactions}\\\vspace{20pt}
\text{\fontsize{15}{18}\selectfont Amir Haluts$^{1}$, Dan Gorbonos$^{2}$, and Nir S. Gov$^{1}$}
\\\vspace{20pt}
\text{\fontsize{11}{15}\selectfont $^{1}$Department of Chemical and Biological Physics, Weizmann Institute of Science, Rehovot 7610001, Israel}
\\\vspace{5pt}
\text{\fontsize{11}{15}\selectfont $^{2}$Department of Collective Behaviour, Max Planck Institute of Animal Behavior, Konstanz 78464, Germany}
    \end{center}
    \setstretch{1.5}
\section*{Abstract}
Active particle systems of interacting self-propelled particles offer a versatile framework for modeling complex systems. When employed to describe aspects of animal behavior, the complexity of animal movement and decision-making often requires the use of unique types of effective interactions between the particles---notably nonreciprocal effective forces that do not obey the usual conservation laws of Newtonian mechanics. Here we review two recent empirically-motivated models, of two very different types of animal behavior, where the behavior is described in terms of active particles which interact through nonreciprocal effective forces. The first model describes the dynamics of animal contests, wherein typically two rivals fight over a localized resource. The uniquely shaped effective potentials between the model's 'contestant particles' manifest the adversarial nature of contest interactions and capture the dynamical essence of contest behavior in space and time. The second model describes the stabilization of cohesive swarms through long-range and adaptive gravity-like attraction. This 'adaptive gravity' model explains the observed mass and velocity profiles of laboratory midge swarms. These examples demonstrate that theoretical models that use the framework of active particles to describe animal behavior can expand the scope of active-particle research, as well as explain complex phenomena in animal behavior.
\end{titlepage}

\clearpage

\setstretch{1.5}
\section*{Introduction}

Active particle systems are composed of interacting, self-propelled particles. In theoretical models, implemented using computer simulations, interactions between particles can take a variety of forms \cite{zottl2023modeling}. These interactions can take the form of the (reciprocal) forces that arise from the potentials between physical particles, for example of the Lennard-Johns-type, but they could also be non-physical, such as those leading to velocity alignment between neighboring particles in Vicsek-type models \cite{vicsek2012collective}. The latter types of interactions are motivated by biological systems, and are often nonreciprocal, for example due to alignment  \cite{motsch2011new,haskovec2013flocking} or due to velocity-aligned anisotropic interaction potentials \cite{barberis2019phase}. The resulting behavior of such non-equilibrium systems is determined by the self-propulsion properties of the particles, the type of interactions, and noise.

Since animals are self-propelled, modeling the movement-related behaviors of animals as individuals and in groups is naturally approached by active particle models \cite{bellomo2022life}. In these systems, the interactions between the animals, which control their movements, are not easy to extract from experimental data. This difficulty stems from these interactions being driven by animal communication, which can involve a complex set of signals, in addition to each animal's interpretation of the movements and intentions of others. The mapping between the biological system and an active particle system is therefore non-trivial. 

Here we review two recent models of animal behavior that describe the dynamics in their respective systems in terms of interacting active particles. These models demonstrate that the effective interactions which arise within the context of animal behavior can have distinct properties that do not appear in typical physical systems: forces between animals can be nonreciprocal, and adaptive.

Nonreciprocal interactions have garnered increasing attention in recent research \cite{durve2018active,saha2020scalar,fruchart2021non,kreienkamp2022clustering,knevzevic2022collective,dinelli2023non}. They arise naturally in the context of animal contests (short-range fights), which we discuss in part \ref{part1} of this chapter. We review a general spatio-temporal model that describes the dynamics of contestants as they escalate their interaction. In the model, the contestants are active particles which interact through effective interaction potentials. This model was originally developed to describe a system of male spider contestants which compete on the two-dimensional orb-web of the female \cite{haluts2021spatiotemporal}, and subsequently extended to describe contests between animals in general \cite{haluts2023modelling}. The dynamics of these agonistic interactions is driven by each contestant's assessment of its own and its rival's 'strength', and the response to these assessments \cite{haluts2023modelling}. The asymmetric scenario where the animals have different 'strengths' therefore naturally gives rise to nonreciprocal interactions, which can be described by each contestant responding to a different effective potential landscape. The most noticeable effect of these nonreciprocal forces during contests is the transition from bounded tussle dynamics (of closely matched rivals) to chase dynamics (between highly unmatched rivals).

In part \ref{part2} of this chapter, we demonstrate the role of 'adaptivity' in the effective interactions between animals. Adaptivity is the property of animals to respond to the relative fold-change in a signal that reaches their sensory organs \cite{adler2018fold}. This means that a constant persistent change in the background signal does not have a lasting effect on the behavior, as the sensory modality adapts to the new background. This property appears in all animals, from bacteria to humans, and in all sensory systems. We review the modeling of adaptive interactions in the context of midge swarms, which form cohesive yet disordered swarms of flying insects \cite{ouellette2022physics}. The model explores long-range attractive interactions between particles, which are adaptive, resulting in an 'adaptive gravity' form \cite{gorbonos2016long,PhysRevE.95.042405,PhysRevResearch.2.013271,gorbonos2020pair}. The adaptivity introduces a non-equilibrium property, by violating the conservation of energy and momentum of the particles, and giving rise to nonreciprocal interactions. Note that we focus here on adaptivity of the animal's sensory sensitivity (and the effective interactions associated with it) due to an external signal, over short timescales that are relevant for the animals' motion, as in adaptive chemotaxis \cite{shklarsh2011smart}. We therefore do not refer to adaptivity on the timescale of genetic mutations, as notably occurs in growing bacterial systems \cite{ben1992adaptive}.

In both parts of this chapter we demonstrate the novel theoretical aspects of the biologically inspired interactions between the particles, as well as compare to experimental data. This chapter emphasizes how modeling animal behavior provides us with motivation to introduce and explore unique types of interactions and emergent phenomena in the context of active particle systems.

\clearpage

\section{Nonreciprocally interacting particles model of animal contests}\label{part1}

It is commonplace for animals to engage in agonistic behavior to obtain and defend food, mates, and territory \cite{briffa2013introduction}. These fights over resources, widely termed 'animal contests', inherently entail the resolution of an interplay between risk and reward. Animal contests have therefore attracted extensive research from ecologists and biologists who sought to understand the evolutionary logic behind such non-trivial trade-offs \cite{payne1996escalation,mesterton1996wars,enquist1983evolution,enquist1990test,parker1981role,hammerstein1982asymmetric,payne1997animals,payne1998gradually,kokko2013dyadic}. From a theoretical standpoint, these risk-reward trade-offs have also made animal contests particularly amenable to the modeling tools of game theory, which, ever since the seminal works of the 1970s \cite{smith1973logic,smith1974theory,parker1974assessment,smith1976logic,parker1979sexual}, has been the central analytical framework used to model and understand them. But while game-theoretic contest models make predictions that can, in principle, be tested empirically \cite{arnott2009assessment}, mapping these models to real contest scenarios has proven difficult, and often lead to ambiguous conclusions \cite{taylor2003mismeasure,arnott2009assessment,pinto2019all,chapin2019further}. An intrinsic hurdle for such mapping has been the omission of within-contest dynamics by game-theoretic models, which typically meet the observable dynamics of contest behavior only in their endpoint predictions (such as who won and how long was the contest), and disregard the spatio-temporal intricacies of the contestants' trajectories. 

While detailed tracking of spatio-temporal trajectories was out of reach for the original contest theorists, it is now a standard practice in animal behavior research \cite{dell2014automated,nathan2022big}. This motivated the development of a new theoretical framework for the dynamics of animal contests, which describes them based on their spatio-temporal properties \cite{haluts2021spatiotemporal,haluts2023modelling}. In this physics-inspired framework, the real-space dynamics of contestants is explicitly described in terms of (active) 'contestant particles', which interact through effective interaction potentials. The adversarial nature of the inter-contestant interactions makes their properties distinct from those of the typical pairwise potentials that govern the interactions between physical particles. Notably, since each contestant's behavior is ultimately driven by the assessment of its own and its rival's 'strength', contest interactions are directional and nonreciprocal. This is handled by assigning the dynamics of each contestant with the influence of a distinct potential landscape that cannot be simply deduced from a global potential. In the following, we discuss the motivation behind this framework, its construction, and some of its implications---both in general and in a specific experimental system. We review the model presented in refs. \cite{haluts2021spatiotemporal} and \cite{haluts2023modelling} as a cohesive body of work, to which we add retrospective insights.

\begin{table}[h!]
  \begin{center}
  \begin{tabular}{m{0.7cm} m{2.5cm} m{9.5cm} m{2.7cm}}
    \toprule
    \textbf{Eq.} & \textbf{Parameter} & \textbf{Role} & \textbf{Value used} \\
    \eqref{eqLangevin} & $\eta$ & Mobility. Sets contribution of deterministic effective forces to the motion of contestant particles & 1.0\hspace{0.2cm}$\left[\mathrm{L\cdot F^{-1}\,T^{-1}}\right]$ \\[1.5em]

    \eqref{eqLangevin} & $D$ & Diffusivity. Sets contribution of stochastic effective forces to the motion of contestant particles & 0.5\hspace{0.2cm}$\left[\mathrm{L^2\,T^{-1}}\right]$ \\[1.5em]

    \eqref{eqVresIsotropic} & $p_i$ & Sets attractiveness of resource, as perceived by contestant $i$ & 4.0\hspace{0.2cm}$\left[\mathrm{F\cdot L}\right]$ \\[1.5em]

    \eqref{eqVresIsotropic} & $\varepsilon$ & Prevents divergence of Logarithmic resource potential & 1.0 \\[1.5em]

    \eqref{eqVjiModel} & $\alpha_{j\rightarrow i}$ & Reflects the motivation of contestant $i$ to escalate the
interaction with the rival $j$ (effective attraction of $i$ towards $j$) & 7.0\hspace{0.2cm}$\left[\mathrm{F\cdot L}\right]$ \\[1.5em]

    \eqref{eqVjiModel} & $\beta$ & Determines the range of effective attraction set by $\alpha_{j\rightarrow i}$ & 1.0 \\[1.5em]

    \eqref{eqVjiModel} & $\delta_{j\rightarrow i}$ & Reflects how intimidating the rival $j$ is perceived by contestant $i$ (effective repulsion of $i$ from $j$) & 3.0\hspace{0.2cm}$\left[\mathrm{F\cdot L}\right]$ \\[1.5em]

    \eqref{eqVjiModel} & $x_0$ & Typical length scale for inter-contestant interaction & 1.0\hspace{0.2cm}$\left[\mathrm{L}\right]$ \\[1.5em]

    \eqref{eqAssessmentFunction} & $m_i$ & Effective 'size' (or 'strength') of contestant $i$ & 0.8--1.2 
    \\[1.5em]

    \eqref{eqAssessmentFunction} & $\alpha_0$ & Sets the basal value of $\alpha_{j\rightarrow i}$ & 7.0\hspace{0.2cm}$\left[\mathrm{F\cdot L}\right]$ 
    \\[1.5em]

    \eqref{eqAssessmentFunction} & $\delta_0$ & Sets the basal value of $\delta_{j\rightarrow i}$ & 3.0\hspace{0.2cm}$\left[\mathrm{F\cdot L}\right]$ 
    \\[1.5em]

    \eqref{eqAssessmentFunction} & $s$ & Power-law exponent representing absolute self-assessment & 1\hspace{0.2cm}[in Eq. \eqref{eqPureSelfAssessment}] \\[1.5em]

    \eqref{eqAssessmentFunction} & $r$ & Power-law exponent representing absolute rival assessment & 1\hspace{0.2cm}[in Eq. \eqref{eqAlternativeMutualAssessment}]
    \\[1.5em]

    \eqref{eqAssessmentFunction} & $s_Q$ & Power-law exponent representing relative self-assessment & 1\hspace{0.2cm}[in Eq. \eqref{eqPureMutualAssessment}]
    \\[1.5em]

    \eqref{eqAssessmentFunction} & $r_Q$ & Power-law exponent representing relative rival assessment & 2\hspace{0.2cm}[in Eq. \eqref{eqPureMutualAssessment}]
    \\[1.5em]

    \eqref{eqCosts} & $K_\mathrm{self}$ & Rate at which 'self-inflicted' costs are accrued & 0.01$\left[\mathrm{T^{-1}}\right]$
    \\[1.5em]

    \eqref{eqCosts} & $K_{ij}$ & Rate at which 'rival-inflicted' costs are accrued & 0.1\hspace{0.2cm}$\left[\mathrm{T^{-1}}\right]$
    \\[0.5em]
    \bottomrule

  \end{tabular}
  \end{center}
  \caption{\textbf{Model parameters.} A list of all parameters used in the model's equations. The column '\textbf{Eq.}' indicates the equation in which each parameter is first used. The column '\textbf{Role}' provides a brief description of each parameter's role in the model. The column '\textbf{Value used}' indicates the numerical value (or range of values) assigned to each parameter, in all or in most cases, and the parameter's dimensions ($\mathrm{L = Length}$, $\mathrm{F = Force}$, and $\mathrm{T = Time}$).}
  \label{tabelParameters}
\end{table}

\subsection{Summary}\label{subsecModelSummary}
The ubiquitous scenario that motivates our modeling approach is that of a dyadic contest, in which two contestants compete over a localized resource. In this generic setup, their mutual attraction to the resource brings the contestants into contest range, and once within this range, they typically become engaged in a short-range contest interaction until they resolve the encounter (disengage)---as illustrated by the trajectories in Fig. \ref{figNonreciprocityAndContest}\textit{C}.

The fundamental assumption behind our theoretical treatment is that the spatio-temporal properties of contest behavior can be effectively described by some archetypal interaction potentials (which give rise to effective interaction forces), and that these potentials predominantly depend on the contestants' positions. Each contestant is treated as an active Brownian particle that moves due to  
deterministic (interaction-induced) and stochastic effective forces---so that its spatio-temporal dynamics is governed by a Langevin equation, as described in section \ref{subsecDynamics}. The attraction towards the resource is encoded by an effective 'resource potential', that can be thought of as the effective potential landscape on which (and due to which) contests take place. The inter-contestant interactions, including the ’contest’ itself, are encoded by effective 'contestant interaction potentials'---the model’s most essential building block, which are constructed based on generic features of contest behavior. These potentials are described in section \ref{subsecEffectivePotentials}.

This framework enables a well-defined notion of a pairwise 'contest' according to the relative motion between the contestants, as described in section \ref{subsecDefinitionPairwiseContest}. Furthermore, the framework can account for different rival assessment strategies, which have been extensively studied in game-theoretic contest models, as described in section \ref{subsecAssessmentStrategies}. The detailed description of the contestants' motion accounts for spatio-temporal features of asymmetric contests. Notably, our model’s contestants exhibit chase dynamics as an emergent property of broken
symmetry in their nonreciprocal interaction forces, as described in \ref{subsecChase}. Finally, explicit time dependence of our model's interaction potentials (for example due to accumulation of fighting costs) is discussed briefly in section \ref{subsecTimeDependence}.

\subsection{Dynamics of contestant particles}\label{subsecDynamics}

To describe the spatio-temporal dynamics of our model's contestants---notably in order to simulate a dyadic contest over a localized resource (as visualized in Fig. \ref{figNonreciprocityAndContest}\textit{C})---we treat each contestant as an active Brownian particle that moves due to deterministic (interaction-induced) and stochastic effective forces. Although the motion characteristics of real animals are typically much richer than those of active particles, such models have proven useful in capturing average properties of animal motion across species, contexts, and timescales \cite{giardina2008collective}, and here we will demonstrate this usefulness for modeling contests. The spatio-temporal dynamics of a 'contestant particle' $i$ (where $i \in \{1,\,2,\,...,\,N\}$ in a system of $N$ contestants) is modeled to be governed by the following (overdamped) Langevin equation (see also \cite{romanczuk2012active} for a comprehensive review of such models for active particle dynamics)
\begin{equation}\label{eqLangevin}
    \dot{\mathbf{r}}_i = -\eta \nabla V_{\mathrm{tot}\rightarrow i}(\mathbf{r}_i,\,\{\mathbf{r}_j\}) + \sqrt{2D}\,\boldsymbol{\xi}_i + \text{activity}.
\end{equation}

Note that we denote vectors by boldface letters and scalars by regular letters throughout this part of the chapter. On the right-hand side  of Eq. \eqref{eqLangevin}, the first term accounts for the deterministic driving force felt by contestant $i$ at position $\mathbf{r}_i$ due to the (total) effective potential $V_{\mathrm{tot}\rightarrow i}$, where $\eta$ is the contestant's 'mobility'. $V_{\mathrm{tot}\rightarrow i}$ generally depends on $\mathbf{r}_i$ and on the positions $\{\mathbf{r}_j\}$ of all other rival contestants, where these positions are defined with respect to a single resource at the origin (as illustrated in Fig. \ref{figResourcePotentials}\textit{A}). The construction of $V_{\mathrm{tot}\rightarrow i}$, which combines all the fundamental building blocks of the model, is detailed in the next section; The second term is the usual translational noise term, where $D$ is an effective translational diffusion coefficient, and $\boldsymbol{\xi}_i$ is a vector of mutually uncorrelated sources of standard Gaussian white noise for each of the spatial dimensions (i.e., $\boldsymbol{\xi}_i = (\xi_{x_i},\,\xi_{y_i})$, with $\xi_{x_i},\,\xi_{y_i}\sim\mathcal{N}(0, 1)$, in two dimensions); The third term accounts for the internal (active) directional persistence of the contestant (which is independent from the directional persistence driven by $V_{\mathrm{tot}\rightarrow i}$) and is commonly governed by its own Langevin equation. While the activity term can be used to better match motion characteristics of particular animal contestants, such as the male spiders studied in ref. \cite{haluts2021spatiotemporal}, for timescales sufficiently longer than the persistence time it mostly acts as an additional source of noise \cite{volpe2014simulation}.

In general, the noise in the model manifests the contribution of 'behavioral noise'---which is integral to biological decision-making---to the motion of the contestants. The practical role of this noise is to allow our simulated contestant particles to dynamically explore the effective potential landscape of $V_{\mathrm{tot}\rightarrow i}$. Notably, this allows us to simulate the fast engage-and-disengage dynamics of typical animal contests \cite{briffa2013introduction,haluts2021spatiotemporal}, as in Fig. \ref{figNonreciprocityAndContest}\textit{C}. In what follows, 
we assume for simplicity that the behavioral properties which make a contestant 'active' can be sufficiently encoded into $V_{\mathrm{tot}\rightarrow i}$, and drop the activity term in Eq. \eqref{eqLangevin}---and thereby also leave the translational diffusion term (driven by Brownian, thermal noise) to solely account for the noise. For the practical implementation of Eq. \eqref{eqLangevin} to simulate the dynamics of interacting contestant particles, yielding contest trajectories as in Figs. \ref{figNonreciprocityAndContest} and \ref{figChase}, we used the first-order (Euler) integration scheme---as detailed in the supporting information of ref. \cite{haluts2023modelling}.

\subsection{Effective interaction potentials and nonreciprocity}\label{subsecEffectivePotentials}
The total effective potential felt by contestant $i$ combines the influence of a resource and and of all other contestants $\{j\}$ (which compete over this particular resource) on the motion of contestant $i$. Observations suggest that when contestants become strongly engaged in a contest interaction, their attention is predominantly given to their rival(s). This can be modeled by an 'attention switch'---where the contestants' motion is affected by their interaction with the resource (which brings them into contest range) only when they are relatively far apart, and is otherwise dominated by their interactions with each other. Therefore, we write $V_{\mathrm{tot}\rightarrow i}$ as follows
\begin{equation}\label{eqVtotGeneral}
V_{\mathrm{tot}\rightarrow i}(\mathbf{r}_i,\,\{\mathbf{r}_j\}) = \sum_{j\,\ne\,i} V_{j\rightarrow i}(\mathbf{r}_i,\,\mathbf{r}_j) + \theta(\mathrm{x}_{ij} -\mathrm{x}_{ij}^\wedge)\,\,V_{\mathrm{res}\rightarrow i}(\mathbf{r}_i)
\end{equation}
where the  $V_{j\rightarrow i}$'s are the 'contestant interaction potentials'---each encoding the influence of a rival contestant $j$ on the motion of contestant $i$, $V_{\mathrm{res}\rightarrow i}$ is the effective 'resource potential' felt by contestant $i$---which encodes the influence of the resource on the motion of contestant $i$, and $\theta(x)$ is a Heaviside step function---where $\mathrm{x}_{ij} \equiv |\mathbf{r}_i - \mathbf{r}_j|/\mathrm{x}_0$ is the dimensionless (scaled) distance between contestants $i$ and $j$ ($\mathrm{x}_0$ is a typical length scale), and $\mathrm{x}_{ij}^\wedge$ is the 'contest onset' distance between $i$ and $j$ (which is defined in section \ref{subsecDefinitionPairwiseContest}). For now, we will treat these effective potentials as time-independent, but note that they could also vary with (interaction) time---notably due to accumulation of fighting costs and learning---as discussed briefly in section \ref{subsecTimeDependence}.

Before discussing explicit functional forms for $V_{\mathrm{res}\rightarrow i}$ and $V_{j\rightarrow i}$, note that Eqs. \eqref{eqLangevin} and \eqref{eqVtotGeneral} can be used to describe the interactions between any number of contestants that compete over a single resource (where the motion of each contestant is governed by its own Langevin equation), and that one can further generalize Eq. \eqref{eqVtotGeneral} to include the influence of more than one resource (such that the contestants move in a 'field' of resources). Nevertheless, although many-contestant and/or many-resource scenarios exist in nature (a specific many-contestant scenario is discussed in section \ref{subsecChase}), the simplest scenario of a dyadic contest, in which two contestants fight over a resource, is by far the most common and well studied \cite{arnott2009assessment,kokko2013dyadic}. For this reason, and since a 'contest' (in our context) is inherently a pairwise concept, we will mostly deal with the dyadic version of Eq. \eqref{eqVtotGeneral},
\begin{equation}\label{eqVtotPair}
V_{\mathrm{tot}\rightarrow i}(\mathbf{r}_i,\,\mathbf{r}_j) = V_{j\rightarrow i}(\mathbf{r}_i,\,\mathbf{r}_j) + \theta(\mathrm{x}_{ij} -\mathrm{x}_{ij}^\wedge)\,\,V_{\mathrm{res}\rightarrow i}(\mathbf{r}_i),
\end{equation}
where contestants '$i$' and '$j$' are, henceforth, the only two contestants (henceforth, we use the indices '$i$' and '$j$' as the dummy identifiers of the contestants, but note that $N = 2$ contestants for all of the remaining text, equations, and figures). We proceed by describing the spatio-temporal features common to contest dynamics in many animal species, and map these generic rules into the shapes of the effective potentials $V_{\mathrm{res}\rightarrow i}$ and $V_{j\rightarrow i}$. The finer details that these potentials have in specific animals and setups can be extracted empirically from contestant trajectories, as described in \cite{haluts2021spatiotemporal} and \cite{haluts2023modelling}. In the following, we often visualize and apply the model in a two-dimensional space, but note that the model is equally applicable in three dimensions. 

\begin{figure}[h!]
    \centering
    \includegraphics[width = 16.5cm]{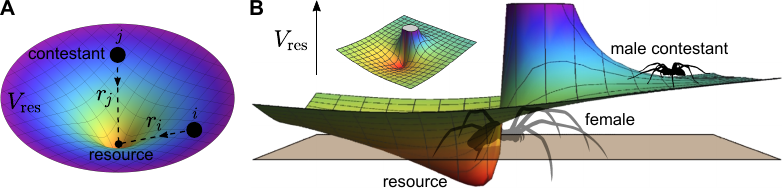}
    \caption{\textbf{Effective resource potentials.} (\textbf{\textit{A}}) The landscape of an effective 'resource potential' with a simple radially-symmetrical form, as in Eq. \eqref{eqVresIsotropic}. (\textbf{\textit{B}}) The landscape of a non-isotropic effective resource potential, which encodes the influence of a female \textit{T. clavipes} spider and its orb-web on the motion of male spider contestants \cite{haluts2021spatiotemporal}.\qquad Adapted from refs. \cite{haluts2023modelling} and \cite{haluts2021spatiotemporal}.}
    \label{figResourcePotentials}
\end{figure}

A typical dyadic contest is initiated when two animals seek access to a spatially localized resource, commonly a territory or a mate \cite{jordan2014reproductive,emlen1977ecology,fuxjager2017animals}. In terms of their physical influence, these resources act as attractors that drive contestants towards each other until conflict becomes inevitable, and can thus be thought of as the effective potential landscape on which (and due to which) contests take place. The shape of this potential landscape depends on the properties of the resource. For most resources, an adequate model might be a simple radially-symmetrical 'sink', for example
\begin{equation}\label{eqVresIsotropic}
    V_{\mathrm{res}\rightarrow i}(\mathbf{r}_i) = p_i \ln(|\mathbf{r}_i| + \epsilon),
\end{equation}
where $p_i,\,\epsilon > 0$ are parameters ($p_i$ sets the attractiveness of the resource, as perceived by contestant $i$). The potential landscape created by Eq. \eqref{eqVresIsotropic} is visualized in Fig. \ref{figResourcePotentials}\textit{A}. For some animate resources, however, behavior and setup could render the shape of the resource potential's landscape less trivial. For example, in the natural contest arena that is the orb-web of a female \textit{Trichonephila clavipes} spider, the resource landscape on which male contestants are moving comprises of the female and the architecture of its web. In this system, males are observed to approach the female from the back and avoid its front---were the potentially cannibalistic female is most reactive  \cite{haluts2021spatiotemporal}. This resource should therefore be modeled by a non-isotropic effective potential. Specifically, in \cite{haluts2021spatiotemporal}, the influence of this resource is described by a potential with a landscape as in Fig. \ref{figResourcePotentials}\textit{B}. Regardless of these system-specific details, the global qualitative effect of resource landscapes is generic: to bring the contestants into contest range due to their mutual attraction to the resource. This generic property implies that effective resource potentials have a global minimum, as in Fig. \ref{figResourcePotentials} \textit{A} and \textit{B}.

The inter-contestant interactions, including the 'contest' itself, are encoded by the contestant interaction potential $V_{j\rightarrow i}$---the model's most essential building block. The archetypal shape of this potential can be constructed based on the following generic (and in a sense minimal) features of contest behavior, which are stated in terms of effective attraction or repulsion between the contestants depending on the inter-contestant distance: (1) Long-range repulsion due to mutual avoidance (which can be surmounted due to the attracting resource), (2) Medium- to short-range attraction when the contestants reach a separation distance in which conflict escalation is inevitable (and hence move towards each other), (3) Strong repulsion at contact, and (4) The strength of the interaction decays to zero when the contestants are far apart. Importantly, conflict escalation is associated with the tendency to decrease the inter-contestant distance (effective attraction), while conflict de-escalation is associated with the tendency to increase this distance (effective repulsion).

\begin{figure}[h!]
    \centering
    \includegraphics[width = 9cm]{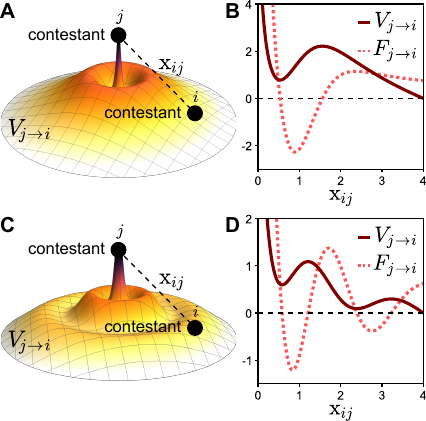}
    \caption{\textbf{Effective contestant interaction potentials.} (\textbf{\textit{A}}) The landscape of an effective 'contestant interaction potential' $V_{j\rightarrow i}$, as given in Eq. \eqref{eqVjiModel}. (\textbf{\textit{B}}) The profiles of $V_{j\rightarrow i}$ and its corresponding force $F_{j\rightarrow i} = -d V_{j\rightarrow i}/d\mathrm{x}_{ij}$ as a function of the inter-contestant distance $\mathrm{x}_{ij}$ (Eq. \eqref{eqVjiModel} with $\alpha_{j\rightarrow i} = 7$, $\delta_{j\rightarrow i} = 3$, and $\beta = 1$). (\textbf{\textit{C}}, \textbf{\textit{D}}) A more elaborate $V_{j\rightarrow i}$ that encodes a two-stage contest escalation scheme.\qquad Adapted from ref. \cite{haluts2023modelling}.}
    \label{figContestantPotentials}
\end{figure}

The above rules amount to a contestant interaction potential with a qualitative shape as in Fig. \ref{figContestantPotentials} \textit{A} and \textit{B}, which can be satisfied by various functional forms. One such particular potential is a combination of a logarithmic repulsion and an attractive Gaussian well,
\begin{equation}\label{eqVjiModel}
    V_{j\rightarrow i}(\mathbf{r}_i,\,\mathbf{r}_j) = - \alpha_{j\rightarrow i}\exp(-\beta{\mathrm{x}_{ij}}^2) - \delta_{j\,\rightarrow \,i}\ln(\mathrm{x}_{ij}),\qquad\mathrm{x}_{ij} = \frac{|\mathbf{r}_i - \mathbf{r}_j|}{\mathrm{x}_0}
\end{equation}
where $\alpha_{j\rightarrow i}$ and $\delta_{j\rightarrow i}$ are positive interaction parameters that set the strengths of effective attraction and repulsion (experienced by contestant $i$ when interacting with a rival $j$), $\beta > 0$ determines the range of effective attraction (assumed to be the same for both contestants), and $\mathrm{x}_0$ is a typical length scale. This potential has two \emph{distinct} local extrema (as in Fig. \ref{figContestantPotentials}\textit{B}), at
\begin{equation}\label{eqExtremaVjiModel}
     \mathrm{x}^\mathrm{min}_{ij} = \sqrt{-\frac{1}{\beta}W_0(\Gamma_{j\rightarrow i})}\quad\text{and}\quad
     \mathrm{x}^\mathrm{max}_{ij} = \sqrt{-\frac{1}{\beta}W_{-1}(\Gamma_{j\rightarrow i})},\quad\text{with}\quad \Gamma_{j\rightarrow i} = -\frac{\delta_{j\rightarrow i}}{2 \alpha_{j\rightarrow i}},
\end{equation} 
(where $W_{-1}(x)$ ($-1/e \leq x < 0$) and $W_0(x)$ ($-1/e \leq x$) are the two real branches of the Lambert $W$ function \cite{corless1996lambert}) provided that
\begin{equation}\label{eqConditionVjiModel}
    \frac{\alpha_{j\rightarrow i}}{\delta_{j\rightarrow i}} > \frac{e}{2},
\end{equation}
where $e$ is Euler's number. The values of $\alpha_{j\rightarrow i}$ and $\delta_{j\rightarrow i}$ are related to contestant $i$'s assessment of itself and of its rival, and therefore depend on the 'strength', or (henceforth) effective 'size', of each contestant. In the animal contest literature, this effective size is typically termed 'resource holding potential' (RHP) \cite{payne1996escalation,mesterton1996wars,enquist1983evolution,enquist1990test,parker1981role,hammerstein1982asymmetric,payne1997animals,payne1998gradually,kokko2013dyadic}. Not to be confused with our model's effective interaction potentials, the RHP is a species-dependent measure for a contestant's ability to win fights, and is commonly correlated (but not strictly interchangeable) with size-related features, such as claw size in crabs or antler size in deer \cite{kokko2013dyadic,arnott2009assessment}. In our case, $\alpha_{j\rightarrow i}$ reflects the motivation of contestant $i$ to escalate the interaction (that is, how aggressive $i$ is) and is therefore associated with the (absolute or relative) effective size of $i$, while $\delta_{j\rightarrow i}$ reflects how intimidating (repulsive) the rival $j$ is perceived by $i$, and is therefore associated with the (absolute or relative) effective size of $j$. Under this interpretation of $\alpha_{j\rightarrow i}$ and $\delta_{j\rightarrow i}$, the case where Eq. \eqref{eqConditionVjiModel} is not satisfied corresponds to contestant $i$ being 'not motivated enough', or equivalently, 'too intimidated', to escalate the interaction with the rival $j$. In section \ref{subsecAssessmentStrategies}, we discuss explicit models for the dependence of $\alpha_{j\rightarrow i}$ and $\delta_{j\rightarrow i}$ on the contestants' effective sizes and the implications of these models in the context of nonreciprocal interactions.

The interaction potential of Eq. \eqref{eqVjiModel} describes a rather minimal contest escalation scheme, in the sense that it features a single escalation barrier (local maximum, see Fig. \ref{figContestantPotentials}\textit{B}) towards a short-range interaction. Proposing this functional form as the 'archetypal' representation of contest interactions is motivated by the premise that contestant dynamics in the vicinity of this ultimate escalation stage account for the qualitative essence of contests in many animal species. Nevertheless, system-specific details that diverge from this minimal description can be included in $V_{j\rightarrow i}$, as long as they correspond to escalation stages that occur at typical inter-contestant distances. For example, such stages exist in contests that escalate through an intermediate evaluation- or display stage \cite{payne1998gradually,popp1987risk,martin2007review,paton1986communication}, where this stage precedes the ultimate escalation into a short-range interaction. These two-stage contest escalation schemes can be encoded by a contestant interaction potential with a qualitative shape as in Fig. \ref{figContestantPotentials} \textit{C} and \textit{D}. In ref. \cite{haluts2021spatiotemporal}, the interaction between male spider contestants---which mostly interact through a single modality of vibratory cues transmitted through the web---was modeled by a similar two-stage potential.

Completing our basic pairwise treatment of contests, a respective interaction potential $V_{i\rightarrow j}$ encodes the influence of contestant $i$ on the motion of the rival $j$. Importantly, since these potentials manifest the (behavioral and perceptual) asymmetries between the contestants, the adversarial interactions encoded by $V_{j\rightarrow i}$ and $V_{i\rightarrow j}$ are nonreciprocal (they violate Newton's third law). This means that, for a general pair of contestants $i$ and $j$, $V_{j\rightarrow i} \ne V_{i\rightarrow j}$ (as illustrated in Fig. \ref{figNonreciprocityAndContest}\textit{A}). The influence of $V_{j\rightarrow i}$ and $V_{i\rightarrow j}$ on the motion of the contestants is governed by the nonreciprocal effective forces that they generate,
\begin{equation}\label{eqNonreciprocalForces}
     \mathbf{F}_{j\rightarrow i}(\mathrm{x}_{ij}) = -\frac{d V_{j\rightarrow i}(\mathrm{x}_{ij})}{d\mathrm{x}_{ij}}\hat{\mathbf{x}}_{ij}\quad\text{and}\quad \mathbf{F}_{i\rightarrow j}(\mathrm{x}_{ij}) = -\frac{d V_{i\rightarrow j}(\mathrm{x}_{ij})}{d\mathrm{x}_{ij}}(-\hat{\mathbf{x}}_{ij}),
\end{equation}
where $\hat{\mathbf{x}}_{ij} = (\mathbf{r}_i - \mathbf{r}_j)/|\mathbf{r}_i - \mathbf{r}_j|$ denotes the unit vector in the $j \rightarrow i$ direction, and again, in general, $\mathbf{F}_{j\rightarrow i} \ne -\mathbf{F}_{i\rightarrow j}$.

\subsection{Definition and duration of a pairwise contest}\label{subsecDefinitionPairwiseContest}

\begin{figure}[h!]
    \centering
    \includegraphics[width = 17.3cm]{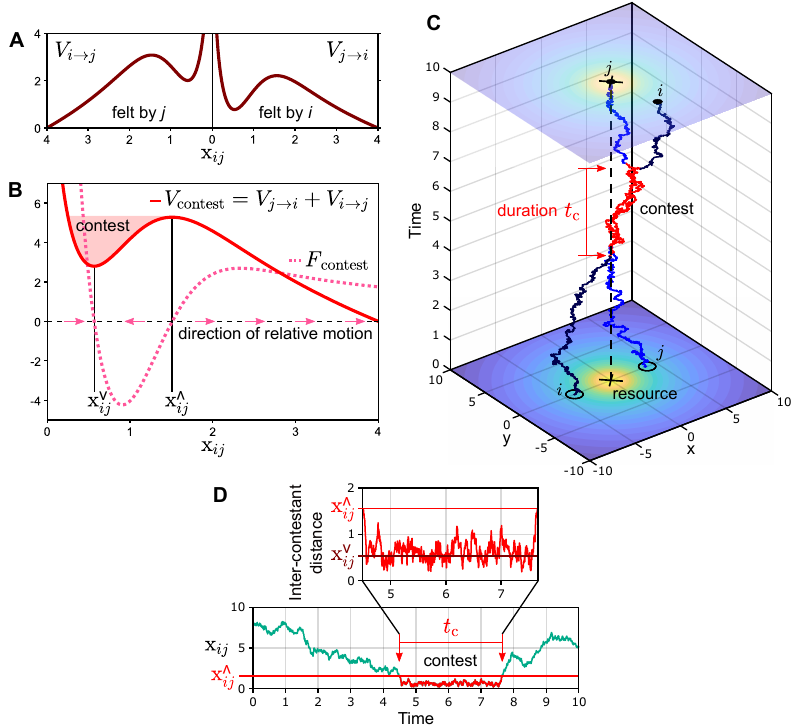}
    \caption{\textbf{Nonreciprocity and definition of a contest.} (\textbf{\textit{A}}) The potentials $V_{j\rightarrow i}$ and $V_{i\rightarrow j}$ encode nonreciprocal interactions that manifest asymmetries between contestants (in general, $V_{j\rightarrow i} \ne V_{i\rightarrow j}$). Here, these asymmetries are described by variations in the interaction parameters of Eq. \eqref{eqVjiModel}, where $V_{j\rightarrow i}$ is shown with $\alpha_{j\rightarrow i} = 7$ and $\delta_{j\rightarrow i} = 3$, and $V_{i\rightarrow j}$ with $\alpha_{j\rightarrow i} = 8$ and $\delta_{j\rightarrow i} = 4$ ($\beta = 1$ for both). (\textbf{\textit{B}}) The relative 'contest potential' $V_\mathrm{contest} = V_{j\rightarrow i} + V_{i\rightarrow j}$ governs the relative motion between the contestants along the inter-contestant axis (arrows). The distance $\mathrm{x}_{ij} = \mathrm{x}_{ij}^\wedge$ is defined as the contest onset. (\textit{\textbf{C}}) Typical simulated trajectories of two identical contestants ($V_{j\rightarrow i} = V_{i\rightarrow j}$) in the vicinity of a resource. The dynamics of each contestant particle was simulated based on Eq. \eqref{eqLangevin}, with the relevant parameter values as in Table \ref{tabelParameters} (see SI of ref. \cite{haluts2023modelling} for further details). Their mutual attraction to the resource brings the contestants into contest range. Segments of the trajectories in which the contestants were engaged in a contest ($\mathrm{x}_{ij} < \mathrm{x}_{ij}^\wedge$) are shown in red. (\textit{\textbf{D}}) The distance between the contestants throughout the simulation. As evident in the close-up view, the contestants spend the majority of the contest near the minimum of $V_\mathrm{contest}$.\qquad
    Adapted from ref. \cite{haluts2023modelling}.}
    \label{figNonreciprocityAndContest}
\end{figure}

While the adversarial inter-contestant interactions that govern contest dynamics are nonreciprocal, to clearly define a 'contest' in our model (in a sense that simultaneously applies to both contestants) it is necessary to define a pairwise contest onset between our model's contestant particles. For this purpose we consider the combined contributions of the effective forces $\mathbf{F}_{j\rightarrow i}$ and $\mathbf{F}_{i\rightarrow j}$ to the relative motion between $i$ and $j$ along the inter-contestant axis. Taking the position of contestant $j$ as a fixed point of reference, contestant $i$ appears to be driven along $\hat{\mathbf{x}}_{ij}$ by a relative 'contest force' $\mathbf{F}_\mathrm{contest}$, given by
\begin{equation}\label{eqFcontest}
    \mathbf{F}_\mathrm{contest}(\mathrm{x}_{ij}) = \mathbf{F}_{j\rightarrow i}(\mathrm{x}_{ij}) - \mathbf{F}_{i\rightarrow j}(\mathrm{x}_{ij}) = -\frac{d}{d \mathrm{x}_{ij}}\left[V_{j\rightarrow i}(\mathrm{x}_{ij}) + V_{i\rightarrow j}(\mathrm{x}_{ij})\right]\hat{\mathbf{x}}_{ij}
\end{equation}
where note that $\mathbf{F}_{i\rightarrow j}$, which is applied by $i$ on $j$ in the $-\hat{\mathbf{x}}_{ij}$ direction, appears to drive $i$ relative to $j$ in the opposing $+\hat{\mathbf{x}}_{ij}$ direction. Eq. \eqref{eqFcontest}, which describes an effectively reciprocal force (it acts on $i$ and $j$ with the same magnitude and in opposing directions), motivates the definition of a corresponding relative 'contest potential' $V_\mathrm{contest}$ as the sum of the individual interaction potentials,
\begin{equation}\label{eqVcontest}
    V_\mathrm{contest}(\mathrm{x}_{ij}) = V_{j\rightarrow i}(\mathrm{x}_{ij}) + V_{i\rightarrow j}(\mathrm{x}_{ij})
\end{equation}
such that $F_\mathrm{contest} = -d V_\mathrm{contest}/d\mathrm{x}_{ij}$. Note that according to the particular contestant interaction potential of Eq. \eqref{eqVjiModel}, the relative contest potential is given by
\begin{equation}\label{eqVcontestModel}
     V_\mathrm{contest}(\mathrm{x}_{ij}) = - (\alpha_{j\rightarrow i} + \alpha_{i\rightarrow j})\exp(-\beta{\mathrm{x}_{ij}}^2) - (\delta_{j\rightarrow i} + \delta_{i\rightarrow j})\ln(\mathrm{x}_{ij}),
\end{equation}
and its two local extrema are at
\begin{equation}\label{eqExtremaVcontestModel}
     \mathrm{x}_{ij}^\vee = \sqrt{-\frac{1}{\beta}W_{0}(\Gamma_{ij})}\quad\text{and}
     \quad\mathrm{x}_{ij}^\wedge = \sqrt{-\frac{1}{\beta}W_{-1}(\Gamma_{ij})},\quad\text{with}\quad\Gamma_{ij} = -\frac{\delta_{j\rightarrow i} + \delta_{i\rightarrow j}}{2\left(\alpha_{j\rightarrow i} + \alpha_{i\rightarrow j}\right)},
\end{equation}
where $\mathrm{x}_{ij}^\vee$ and $\mathrm{x}_{ij}^\wedge$ denote the locations of the minimum and maximum, respectively, as indicated in Fig. \ref{figNonreciprocityAndContest}\textit{B}. The contest potential of Eq. \eqref{eqVcontestModel} has these two \emph{distinct} extrema when
\begin{equation}\label{eqConditionVcontestModel}
    \frac{\alpha_{j\rightarrow i} + \alpha_{i\rightarrow j}}{\delta_{j\rightarrow i} + \delta_{i\rightarrow j}} > \frac{e}{2},
\end{equation}
and becomes repulsive for all $\mathrm{x}_{ij}$ when this inequality is not satisfied.

Regardless of the particular functional form of Eq. \eqref{eqVcontestModel}, but assuming that $V_\mathrm{contest}$ has a qualitative shape as in Fig. \ref{figNonreciprocityAndContest}\textit{B} (with corresponding local extrema at some $\mathrm{x}_{ij}^\vee$ and $\mathrm{x}_{ij}^\wedge$), we now define the onset of a contest according to the direction of relative motion between the contestants---governed by the sign of $F_\mathrm{contest}$ (as illustrated in Fig. \ref{figNonreciprocityAndContest}\textit{B}). When the contestants reach a separation $\mathrm{x}_{ij}$ that is shorter than $\mathrm{x}_{ij}^\wedge$ (driven by their mutual attraction to the resource and the stochastic components of their motion), the relative force becomes attractive ($F_\mathrm{contest} < 0$) towards $\mathrm{x}_{ij}^\vee$, and the inter-contestant dynamics is characterized by a transient bounded state. In Fig. \ref{figNonreciprocityAndContest} \textit{C} and \textit{D}, the dynamics of two contestant particles getting in and out of this transient bounded state is illustrated by simulated contestant trajectories. We identify this bounded state with the ultimate escalation into a short-range contest, in which the contestants are completely engaged with each other, and define $\mathrm{x}_{ij}^\wedge$ as the contest onset distance---such that two of our model's contestant particles are considered to be 'engaged in a contest' when $\mathrm{x}_{ij} < \mathrm{x}_{ij}^\wedge$ (in Eqs. \eqref{eqVtotGeneral} and \eqref{eqVtotPair}, this is the regime where $\theta = 0$). Note that even if the contest potential has a more elaborate shape than in Fig. \ref{figNonreciprocityAndContest}\textit{B} (as in Fig. \ref{figContestantPotentials}\textit{D} and ref. \cite{haluts2021spatiotemporal}), the above principles can still be used to define an analogous pairwise contest onset---notably by taking the onset of the most short-range potential well to play the role of $\mathrm{x}_{ij}^\wedge$.

From our definition of a contest, it immediately follows that contest duration ($t_\mathrm{c}$)---a common measurable of interest in studies of animal contests \cite{arnott2009assessment}---is equivalent to the escape time of our contestant particles from their transient bounded state within the potential well of $V_\mathrm{contest}$
(as illustrated in Fig. \ref{figNonreciprocityAndContest}\textit{D}). Then, with an effective 'contest bounding energy' of
\begin{equation}
    \label{eqContestBoundingEnergy}
    U = V_\mathrm{contest}(\mathrm{x}_{ij}^\wedge) - V_\mathrm{contest}(\mathrm{x}_{ij}^\vee),
\end{equation}
the mean contest duration is analogous to the Karmer's escape time \cite{hanggi1990reaction}, and thus follows an Arrhenius' law 
\begin{equation}
    \label{eqContestDuration}
    \langle t_\mathrm{c} \rangle \simeq \Pi\,\exp\left(\frac{U}{T_\mathrm{eff}}\right),
\end{equation}
where the pre-exponential factor $\Pi$ mostly depends on the potential's curvature at its extrema, and the effective temperature $T_\mathrm{eff}$ mostly depends on the properties of the active particle---namely on the diffusivity $D$ and the activity \cite{hanggi1990reaction,geiseler2016kramers}. Eq. \eqref{eqContestDuration} assumes that, after they first cross the contest onset $\mathrm{x}_{ij}^\wedge$, the contestants reach the minimum $\mathrm{x}_{ij}^\vee$ quickly---within an average time that is significantly shorter than $t_\mathrm{c}$---and spend the rest of the contest near $\mathrm{x}_{ij}^\vee$ until they escape, as in Fig. \ref{figNonreciprocityAndContest}\textit{D}. Note that this implies that both the translational diffusion and the activity terms in Eq. \eqref{eqLangevin} should be taken to be sufficiently small compared to the contest bounding energy $U$. Deriving analytical expressions for $\Pi$ and $T_\mathrm{eff}$ in the context of our active particles setup is possible in some limits (see \cite{geiseler2016kramers} for a rather instructive derivation for active particles obeying the same Langevin Equation as ours). But importantly, one fundamental assumption in Kramer's analytical derivation \cite{hanggi1990reaction}, that the potential well is very deep compared to the stochastic fluctuations, is generally not satisfied for our contestant particles---as they have to describe the fast engage-and-disengage dynamics of typical animal contests \cite{briffa2013introduction,haluts2021spatiotemporal}. Nevertheless, even without exact expressions for $\Pi$ and $T_\mathrm{eff}$,
Eq. \eqref{eqContestDuration} can be used to predict the qualitative trends of $\langle t_\mathrm{c} \rangle$ with respect to variations in the parameters of $V_{j\rightarrow i}$ and $V_{i\rightarrow j}$ (which determine $U$), as shown in ref. \cite{haluts2023modelling}.

\subsection{Assessment strategies and interaction asymmetries}\label{subsecAssessmentStrategies}
The type of RHP-related information  used by contestants in their decision-making, commonly termed the 'assessment strategy', has been extensively studied in game-theoretic contest models---especially in order to describe asymmetric contests \cite{arnott2008information,kokko2013dyadic,smith1976logic} [recall the definition of RHP, given after Eq. \eqref{eqConditionVjiModel}]. In \cite{haluts2023modelling}, we proposed that different assessment strategies can be expressed in our framework in terms of an 'assessment function' $\mathbf{A}_{j\rightarrow i}$ (of contestant $i$ with respect to a rival $j$), which defines the functional relations between the interaction parameters of $V_{j\rightarrow i}$ and $V_{i\rightarrow j}$ and the contestants' effective sizes---where the effective size of contestant $i$ is defined as a dimensionless proxy for its RHP,
\begin{equation}\label{eqEffectiveSizeDefinition}
    m_i = \frac{\text{RHP of contestant $i$}}{\text{RHP of reference}},
\end{equation}
with the 'RHP of reference' being, for example, the population average. For the contestant interaction potential $V_{j\rightarrow i}$ of Eq. \eqref{eqVjiModel}, with interaction parameters $\alpha_{j\rightarrow i}$ and $\delta_{j\rightarrow i}$, we write $\mathbf{A}_{j\rightarrow i}$ as
\begin{equation}\label{eqAssessmentFunction}
    \mathbf{A}_{j\rightarrow i}(m_i,\,m_j) \equiv
    \begin{pmatrix}
    \alpha_{j\rightarrow i}(m_i,\,m_j) \\ 
    \delta_{j\rightarrow i}(m_i,\,m_j) 
    \end{pmatrix} = 
    \begin{pmatrix}
    \alpha_0\,{m_i}^s(m_i/m_j)^{s_Q} \\ 
    \delta_0\,{m_j}^r(m_j/m_i)^{r_Q} 
    \end{pmatrix},
\end{equation}
where $\alpha_0$ and $\delta_0$ are scaling parameters, and the different power laws represent: ${m_i}^s$---absolute self-assessment; ${m_j}^r$---absolute rival-assessment; $(m_i/m_j)^{s_Q}$ and $(m_j/m_i)^{r_Q}$---relative self- and rival-assessment, respectively. These relations are best understood by recalling that $\alpha_{j\rightarrow i}$ reflects the motivation of contestant $i$ to escalate the interaction with the rival $j$, and therefore increases with $m_i$ but decreases with $m_j$, while $\delta_{j\rightarrow i}$ reflects how intimidating the rival $j$ is perceived by $i$, and therefore decreases with $m_i$ but increases with $m_j$. Note that the rival's assessment function, $\mathbf{A}_{i\rightarrow j} = \left(\alpha_{i\rightarrow j},\,\delta_{i\rightarrow j}\right)^T$, is obtained by swapping the functional roles of $m_i$ and $m_j$ in Eq. \eqref{eqAssessmentFunction} [that is, $\mathbf{A}_{i\rightarrow j}(m_i,\,m_j) = \mathbf{A}_{j\rightarrow i}(m_j,\,m_i)$]. The assessment strategy, which is now defined in terms of how $\alpha_{j\rightarrow i}$ and $\delta_{j\rightarrow i}$ vary with $m_i$ and $m_j$, is represented by the (non-negative) values of the exponents $s$, $r$, $s_Q$, $r_Q$. In particular, with $s > 0$ and $r,\,s_Q,\,r_Q = 0$, we obtain our model's representation of the so-called 'pure self-assessment' \cite{arnott2009assessment},
\begin{equation}\label{eqPureSelfAssessment}
    \mathbf{A}_{j\rightarrow i}(m_i,\,m_j) =
    \begin{pmatrix}
    \alpha_0\,{m_i}^s \\ 
    \delta_0 
    \end{pmatrix}\quad\text{for pure self-assessment},
\end{equation}
where contestants only assess their own (absolute) effective size and disregard their rival's effective size, and with $s,\,r = 0$ and $s_Q,\,r_Q > 0$, we obtain the representation for pure 'mutual assessment' \cite{arnott2009assessment},
\begin{equation}\label{eqPureMutualAssessment}
    \mathbf{A}_{j\rightarrow i}(m_i,\,m_j) =
    \begin{pmatrix}
    \alpha_0\,(m_i/m_j)^{s_Q} \\ 
    \delta_0\,(m_j/m_i)^{r_Q} 
    \end{pmatrix}\quad\text{for pure mutual assessment},
\end{equation}
where the contestants conduct a purely relative assessment of their own effective size with respect to their rival's effective size. Note that, by construction, for $m_i = m_j = 1$ ($\text{'RHP of contestant $i$'} = \text{'RHP of reference'}$ in Eq. \eqref{eqEffectiveSizeDefinition}), any mode of assessment described by Eq. \eqref{eqAssessmentFunction} yields the same interaction potentials.

\begin{figure}[h!]
    \centering
    \includegraphics[width = 17.3cm]{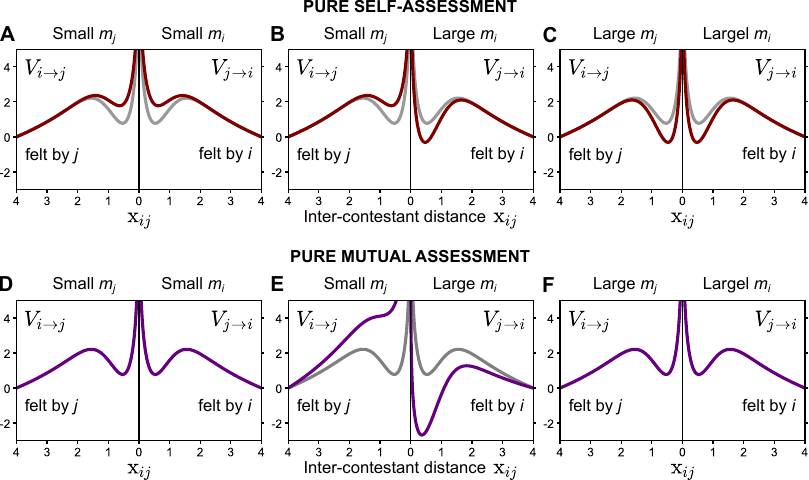}
    \caption{\textbf{Assessment strategies in terms of contestant interaction potentials.}  Under pure self-assessment (Eq. \eqref{eqPureSelfAssessment} with $s = 1$) and pure mutual assessment (Eq. \eqref{eqPureMutualAssessment} with $s_Q = 1$ and $r_Q = 2$), the interaction potentials $V_{i\rightarrow j}$ and $V_{j\rightarrow i}$ are shown for the interaction of (\textbf{\textit{A}}, \textbf{\textit{D}}) small size-matched contestants ($m_j = m_i = 0.8$), (\textbf{\textit{B}}, \textbf{\textit{E}}) a small contestant with a large contestant ($m_j = 0.8,\,m_i = 1.2$ for pure self-assessment, and $m_j = 0.9,\,m_i = 1.1$ for pure mutual assessment [different values merely for graphing purposes]), and (\textbf{\textit{C}}, \textbf{\textit{F}}) large size-matched contestants ($m_j = m_i = 1.2$). For reference, each graph also shows the potentials of medium size-matched contestants ($m_j = m_i = 1$) in grey. Eqs. \eqref{eqPureSelfAssessment} and \eqref{eqPureMutualAssessment} were used with $\alpha_0 = 7$ and $\delta_0 = 3$.\qquad Adapted from ref. \cite{haluts2023modelling}.}
    \label{figAssessment}
\end{figure}

Fig. \ref{figAssessment} illustrates the differences between the assessment functions of Eqs. \eqref{eqPureSelfAssessment} and \eqref{eqPureMutualAssessment} in terms of how $V_{i\rightarrow j}$ and $V_{j\rightarrow i}$ vary with $m_i$ and $m_j$; It shows these potentials, based on Eq. \eqref{eqVjiModel}, under pure self-assessment (Eq. \eqref{eqPureSelfAssessment} with $s = 1$) and pure mutual assessment (Eq. \eqref{eqPureMutualAssessment} with $s_Q = 1$ and $r_Q = 2$), for the interactions of small size-matched contestants, a small contestant with a large contestant, and large size-matched contestants---where 'small' and 'large' refer to the contestants' effective sizes. Notably, compared to pure self-assessment---which detects absolute changes in $m_i$ and $m_j$ (compare Fig. \ref{figAssessment} \textit{A}, \textit{B}, and \textit{C}), pure mutual assessment is scale-invariant, as Eq. \eqref{eqPureMutualAssessment} satisfies $\mathbf{A}_{i\rightarrow j}(c\,m_i,\,c\,m_j) = \mathbf{A}_{i\rightarrow j}(m_i,\,m_j)$ for any scalar $c > 0$. This property is illustrated in Fig. \ref{figAssessment} \textit{D} and \textit{F} by the fact that, under pure mutual assessment, the interaction potentials of two size-matched small contestants and two size-matched large contestants---are identical. In addition, detection of asymmetries between $m_i$ and $m_j$ is much stronger in mutual assessment, as evident by the pronounced broken symmetry between $V_{j\rightarrow i}$ and $V_{i\rightarrow j}$ for the interaction of unmatched contestants (compare Fig. \ref{figAssessment} \textit{B} and \textit{E}).

\begin{figure}[h!]
    \centering
    \includegraphics[width = 17.3cm]{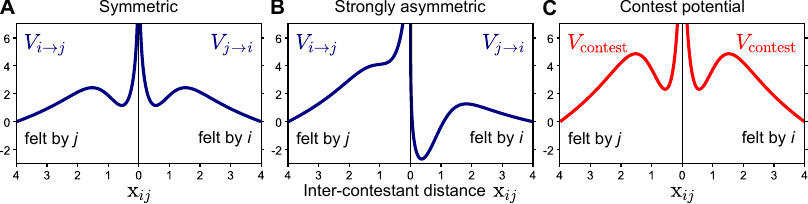}
    \caption{\textbf{Very different contest interactions with the same contest potential.} (\textbf{\textit{A}}) The contestant interaction potentials of a symmetric interaction between two identical contestants ($V_{j\rightarrow i} = V_{i\rightarrow j}$, with $\alpha_{j\rightarrow i} = \alpha_{i\rightarrow j} = 7.14$, $\delta_{j\rightarrow i} = \delta_{i\rightarrow j} = 3.25$, and $\beta = 1$). (\textbf{\textit{B}}) The contestant interaction potentials of a  strongly asymmetric interaction ($\alpha_{j\rightarrow i} = 8.55$, $\alpha_{i\rightarrow j} = 5.73$, $\delta_{j\rightarrow i} = 2.02$, $\delta_{i\rightarrow j} = 4.48$, and $\beta = 1$), in which $V_{i\rightarrow j}$ is strictly repulsive. (\textbf{\textit{C}}) These very different interactions have the same contest potential $V_\mathrm{contest} = V_{j\rightarrow i} + V_{i\rightarrow j}$, since $\alpha_{j\rightarrow i} + \alpha_{i\rightarrow j}$ and $\delta_{j\rightarrow i} + \delta_{i\rightarrow j}$ (and $\beta$) are the same for both.}
    \label{figVeryDifferentInteractions}
\end{figure}

It should be noted that even a highly asymmetric inter-contestant interaction as in Fig. \ref{figAssessment}\textit{E} (where notably, $V_{i\rightarrow j}$ is strictly repulsive while $V_{j\rightarrow i}$ has a deep attractive well with a lowered escalation barrier), can still describe a pairwise 'contest' as defined in section \ref{subsecDefinitionPairwiseContest}---as long as the potential well of $V_\mathrm{contest} = V_{j\rightarrow i} + V_{i\rightarrow j}$ exists (that is, as long as $V_\mathrm{contest}$ supports a transient bounded state for some $\mathrm{x}_{ij} < \mathrm{x}_{ij}^\wedge$). This is illustrated in Fig. \ref{figVeryDifferentInteractions} by very different contest interactions (in terms of the asymmetry between $V_{j\rightarrow i}$ and $V_{i\rightarrow j}$) that were chosen to yield the same contest potential. As discussed in section \ref{subsecChase}, the qualitative properties of contest trajectories with symmetric interactions (as in Fig. \ref{figVeryDifferentInteractions}\textit{A}) will clearly differ from those of contests with strongly asymmetric interactions (as in Fig. \ref{figVeryDifferentInteractions}\textit{B}). But interestingly, the mean contest duration predicted by Eqs. \eqref{eqContestBoundingEnergy} and \eqref{eqContestDuration} is the same for both of these interactions (in \cite{haluts2023modelling}, this property is shown in simulations), as they share the same contest potential of Fig. \ref{figVeryDifferentInteractions}\textit{C}. From a behavioral perspective, this can be understood through the heuristic that both interactions have the same 'total' amounts of motivation ($\alpha_{j\rightarrow i} + \alpha_{i\rightarrow j}$) and intimidation ($\delta_{j\rightarrow i} + \delta_{i\rightarrow j}$), and therefore persist for the same amount of time, although in Fig. \ref{figVeryDifferentInteractions}\textit{A}, these are split equally between the contestants, while in Fig. \ref{figVeryDifferentInteractions}\textit{B}, contestant $i$ has most of the motivation and contestant $j$ has most of the intimidation.

\subsection{Chase dynamics emerges from interaction asymmetry}\label{subsecChase}

\begin{figure}[h!]
    \centering
    \includegraphics[width = 16cm]{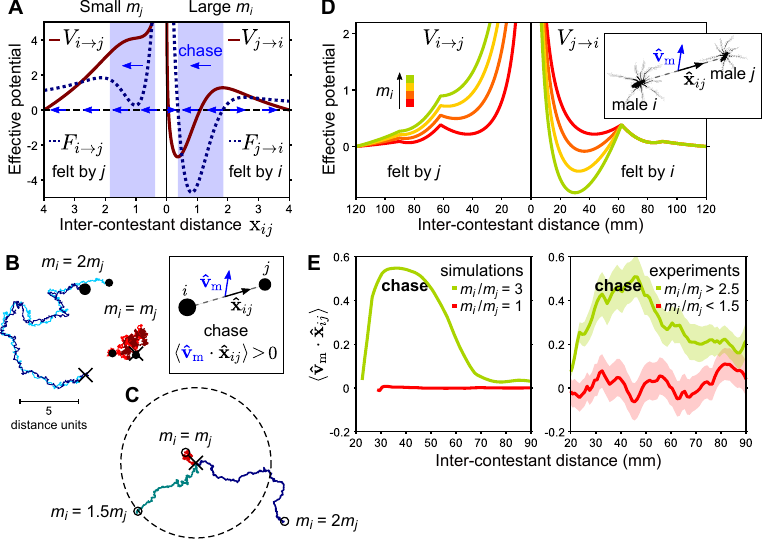}
    \caption{\textbf{Chase dynamics emerges from interaction asymmetry.} (\textit{\textbf{A}}) Strongly asymmetric interaction potentials illustrate how interaction asymmetry leads to chase dynamics. Shaded rectangles mark the chase range, within which the larger contestant is attracted by a deep minimum, while the smaller contestant is repelled in the same direction. Arrows indicate the direction of motion for each contestant, as dictated by the signs of the forces $F_{j\rightarrow i}$ and $F_{i\rightarrow j}$. (\textit{\textbf{B}}) Typical trajectories of strongly asymmetric and symmetric contests in simulations. Chase dynamics is quantified by the direction correlation between the velocity of the contestants' midpoint---measured by $\hat{\mathbf{v}}_\mathrm{m}$, and the inter-contestant direction $\hat{\mathbf{x}}_{ij}$ (see \textit{Inset}). (\textit{\textbf{C}}) Trajectories of the contestants' midpoint during contests under pure mutual assessment (the dependence of $\alpha_{j\rightarrow i}$ and $\delta_{j\rightarrow i}$ on $m_i$ and $m_j$ is given by Eq. \eqref{eqPureMutualAssessment}, with $s_Q = 1$ and $r_Q = 2$) for different effective size ratios. These trajectories, of equal durations, demonstrate the effect of chase dynamics on the contestants' displacement during the contest. In \textit{B} and \textit{C}, 'X' marks the contestants' midpoint at contest onset and circles mark the final positions. (\textit{\textbf{D}}) Increasing asymmetry in the contestant interaction potentials of ref. \cite{haluts2021spatiotemporal} (used to model the interaction between male spider contestants) as a function of increasing $m_i \geq m_j$ (at constant $m_j$). (\textit{\textbf{E}}) Profiles of $\langle\hat{\mathbf{v}}_\mathrm{m}\cdot\hat{\mathbf{x}}_{ij}\rangle$ in contests between two male spiders (from ref. \cite{haluts2021spatiotemporal}), in simulations and experiments, for symmetric and asymmetric contests. In experiments, $m_i$ and $m_j$ denote the actual body weights of the spiders. Error bars show SEM.\qquad Adapted from refs. \cite{haluts2023modelling} and \cite{haluts2021spatiotemporal}. See these works for further details.}
    \label{figChase}
\end{figure}

Chasing behavior is evidently a ubiquitous feature of agonistic interactions in animals, and is intuitively associated with some form of asymmetry between the chasing individual and the one being chased. The essence of this behavior is captured by our model's contestant particles, which exhibit chase dynamics as an emergent property of broken symmetry in their nonreciprocal interaction forces. As illustrated in Fig. \ref{figChase}\textit{A} for strongly asymmetric contestant interaction potentials, within the 'chase' range the larger contestant is attracted by a deep potential well which drives it towards its smaller rival, while the smaller contestant is repelled in the same direction---away from its larger rival. Together, these effects gives rise to directed chase behavior during the contest. This 'microscopic' chase---at the single-particle level---is analogous, both in qualitative nature and in the mechanism that creates it, to the macroscopic chase and traveling wave phenomena that occur between (chemical) species in active mixtures in the presence of nonreciprocal interactions \cite{saha2020scalar,you2020nonreciprocity}.

Fig. \ref{figChase} \textit{B} and \textit{C} compare typical trajectories of strongly asymmetric and symmetric contests under 'pure mutual assessment' in simulations (namely, the dependence of $\alpha_{j\rightarrow i}$ and $\delta_{j\rightarrow i}$ on the contestants' effective sizes is given by Eq. \eqref{eqPureMutualAssessment}, with $s_Q = 1$ and $r_Q = 2$). While symmetric contests are characterized by scrambled and relatively localized trajectories, the trajectories of strongly asymmetric contests feature substantial directional alignment and persistence, and consequently much greater displacement, due to chase dynamics. The extent to which chase dynamics governs the contestants' trajectories can be quantified by considering the direction correlation between the velocity of the contestants' midpoint---measured by $\hat{\mathbf{v}}_\mathrm{m}$, and the inter-contestant direction $\hat{\mathbf{x}}_{ij}$, as defined in Fig. \ref{figChase}\textit{B} (\textit{Inset}). The temporal mean of this 'chase correlator', $\langle \hat{\mathbf{v}}_\mathrm{m}\cdot \hat{\mathbf{x}}_{ij} \rangle$, is zero for symmetric interactions, and $\left|\langle \hat{\mathbf{v}}_\mathrm{m}\cdot \hat{\mathbf{x}}_{ij} \rangle\right| \sim 1$ if the interaction is dominated by the chase phase.

Accounting for the dynamical aspects of asymmetric contests was particularly important in the natural system of \textit{T. clavipes} male spider contestants studied in ref. \cite{haluts2021spatiotemporal}---as mature specimen of these male spiders can vary greatly in size \cite{vollrath1980male,schneider2000sperm}. The observed coexistence of sexually-mature \textit{T. clavipes} males that differ in their weights by as much as an order of magnitude is puzzling, as one would expect stabilizing evolution to prevent such extraordinary size variability \cite{rittschof2010male}. 
Fig. \ref{figChase}\textit{D} shows the experimentally-calibrated, two-stage interaction potentials used to model the interactions between the spider contestants of ref. \cite{haluts2021spatiotemporal}. The observation-based model for assessment in this system (if it were to be mapped to the one-stage interaction potential of Eq. \eqref{eqVjiModel}) corresponds to the following assessment function
\begin{equation}\label{eqAlternativeMutualAssessment}
    \mathbf{A}_{j\rightarrow i}(m_i,\,m_j) = 
     \begin{pmatrix}
    \alpha_0\,m_i \\ 
    \delta_0\,m_j 
    \end{pmatrix}.
\end{equation}
The way in which Eq. \eqref{eqAlternativeMutualAssessment} detects effective size asymmetries is qualitatively similar to the pure mutual assessment of Eq. \eqref{eqPureMutualAssessment}, as evident by how $V_{j\rightarrow i}$ and $V_{i\rightarrow j}$ vary with an increasing $m_i$ in Fig. \ref{figChase}\textit{D}. Accordingly, it predicts substantial chase dynamics for large asymmetries in the contestants' effective sizes, as shown in Fig. \ref{figChase}\textit{E} \textit{Left}. This prediction is in good agreement with the dynamics of real spider contests, as shown in Fig. \ref{figChase}\textit{E} \textit{Right}.

The ability of the model in ref. \cite{haluts2021spatiotemporal} to account for these details motivated its application to a many-contestant scenario taking place in this system. It was found that local male density, defined as the number of males occupying a single female's web, is highly variable in wild populations of \textit{T. clavipes} \cite{rittschof2010male}. Some studies suggested that this could resolve the puzzle of high size variability, if large males have a significant reproductive advantage when male density is high, but this advantage is diminished when male density is low---allowing for a wide distribution of sizes to be stable \cite{rittschof2010male}. Using simulations, we provided in ref. \cite{haluts2021spatiotemporal} a mechanistic explanation for the hypothesized competitive advantage of large males at high male densities. We have shown that the repulsive effective interaction potentials inflicted by large males on smaller males enable them to resolve the interactions with their smaller rivals swiftly and at a distance---and thereby reach the female-resource faster and retain it better, while small males form long-lasting contests with other small males (which are more abundant, \cite{rittschof2010male}), and therefore suffer disproportionately from a high density of rivals, in which the frequency of contests is high. Conversely, when male density is low, we have found in simulations that large males have a less significant advantage at reaching and retaining the female-resource \cite{haluts2021spatiotemporal}. On such female webs, which coexist in nature with high-density webs \cite{rittschof2010male}, the disadvantages of being large (e.g. higher detectability by the potentially cannibalistic female or by other predators), may outweigh its benefits.

\subsection{Time dependence of contest interactions}\label{subsecTimeDependence}
Until now we have treated inter-contestant interactions as if they do not (explicitly) depend on time: the effective potentials that describe these interactions were time-independent, and so the only 'time dependence' in our framework was due to the motion of the contestant particles (driven by their governing Langevin equations), which continuously alters the potential landscapes that they experience. However, contestant interaction potentials reflect the contestants' current effective sizes, and these could vary during (and due to) the interaction---notably as fighting costs accumulate (and decrease the effective sizes compared to their initial states) and as the contestants update their perception of their own and of their rival's status (learning). The dependence of $m_i$ and $m_j$ on interaction time due to such effects introduce into $V_{j\rightarrow i}$ and $V_{i\rightarrow j}$ an explicit time dependence.

\begin{figure}[h!]
    \centering
    \includegraphics[width = 16cm]{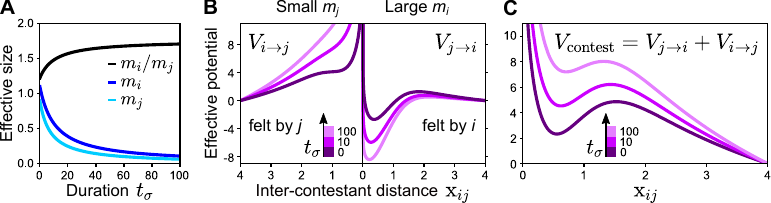}
    \caption{\textbf{Time dependence of contest interactions.} (\textit{\textbf{A}}) The dynamics of the effective sizes $m_i$ and $m_j$, and of their ratio $m_i/m_j$, according to Eq. \eqref{eqCosts} with $K_\mathrm{self} = 0.01$, $K_{ij} = 0.1$, and with initial effective sizes $\mu_i = 1.1$ and $\mu_j = 0.9$. (\textit{\textbf{B}}) The dependence of $m_i$ and $m_j$ on the accumulated contest duration $t_\sigma$ due to costs means that $V_{i\rightarrow j}$ and $V_{j\rightarrow i}$ themselves depend on $t_\sigma$. Here, these potentials vary with the values of $m_i$ and $m_j$ in \textit{A} as $t_\sigma$ increases, according to pure mutual assessment (Eq. \eqref{eqPureMutualAssessment} with $s_Q = 1$ and $r_Q = 2$). (\textit{\textbf{C}}) Under this assessment strategy, the bounding well that defines the contest regime becomes shallower with $m_i/m_j$, and therefore with $t_\sigma$.}
    \label{figTimeDependence}
\end{figure}

As an illustrative example for the notable consequences of such effects on the properties of the inter-contestant interaction, consider the simple model from ref. \cite{haluts2023modelling} for the dependence of the effective sizes on the accumulated contest time $t_\sigma$, given by
\begin{equation}\label{eqCosts}
    m_i(t_\sigma) = \frac{\mu_i}{1 + \left(K_\mathrm{self} + K_{ij}\,\dfrac{\mu_j}{\mu_i}\right)t_\sigma}\qquad\text{and}\qquad m_j(t_\sigma) = \frac{\mu_j}{1 + \left(K_\mathrm{self} + K_{ij}\,\dfrac{\mu_i}{\mu_j}\right)t_\sigma},
\end{equation}
where $\mu_i$ and $\mu_j$ are the effective sizes of contestants $i$ and $j$ when the contest starts, $K_\mathrm{self}$ is the rate at which 'self-inflicted' costs are accrued (e.g., average rate of energy expenditure), $K_{ij}\,\mu_j/\mu_i$ is the rate at which 'rival-inflicted' costs are accrued (e.g., average rate of injuries), and $t_\sigma$ is accumulated only during the contest (that is, when $\mathrm{x}_{ij} < \mathrm{x}_{ij}^\wedge$). Eq. \eqref{eqCosts} naively assumes that costs are accrued continuously and deterministically, and that the rates of accruing them are independent of the instantaneous values of $m_i$ and $m_j$. Nevertheless, it includes a simple form of feedback, since the rate of incurring costs from the rival is proportional to the initial effective size ratio---such that the larger contestant inflicts costs faster but incurs them slower. Note that according to Eq. \eqref{eqCosts}, while both $m_i$ and $m_j$ decrease monotonically with $t_\sigma$, their ratio $m_i/m_j$ increases with $t_\sigma$ when $\mu_i > \mu_j$ and $K_{ij} > 0$. It approaches a constant as $t_\sigma \rightarrow \displaystyle\infty$, where
\begin{equation}\label{eqRatioLimit}
    \lim_{t_\sigma\,\to\,\infty}\frac{m_i}{m_j} = \frac{\mu_i}{\mu_j}\cdot\frac{K_\mathrm{self}\,\mu_j + K_{ij}\,{\mu_i}^2}{K_\mathrm{self}\,\mu_i + K_{ij}\,{\mu_j}^2}.
\end{equation}
The dynamics of $m_i$, $m_j$, and $m_i/m_j$ according to Eq. \eqref{eqCosts} is shown in Fig. \ref{figTimeDependence}\textit{A} for a case where $\mu_i > \mu_j$.

Now, through the assessment function of Eq. \eqref{eqAssessmentFunction}, $V_{j\rightarrow i}$ and $V_{i\rightarrow j}$ also depend on $t_\sigma$. In Fig. \ref{figTimeDependence}\textit{B}, this dependence is visualized under pure mutual assessment (Eq. \eqref{eqPureMutualAssessment} with $s_Q = 1$ and $r_Q = 2$). The increasing asymmetry (in terms of the growing ratio $m_i/m_j$) is manifested by an increasingly asymmetric inter-contestant interaction---as evident by the opposing trends of the potentials with $t_\sigma$ in Fig. \ref{figTimeDependence}\textit{B}. An interesting consequence of this interaction-induced amplification of asymmetry is that it offers a generic mechanism for the emergence of strong chase dynamics during the contest, even if the initial asymmetry is small. Moreover, note that under Eq. \eqref{eqPureMutualAssessment} with $s_Q = 1$ and $r_Q = 2$, the bounding well that defines the contest regime becomes shallower with $m_i/m_j$, and therefore with $t_\sigma$ (as shown in Fig. \ref{figTimeDependence}\textit{C})---which in turn reduces the mean contest duration (recall Eq. \eqref{eqContestDuration}). This trend can eventually lead to a cost-driven contest termination when the bounding well completely disappears and $V_\mathrm{contest}$ becomes strictly repulsive (recall that this happens when Eq. \eqref{eqConditionVcontestModel} is not satisfied), as explored in detail in ref. \cite{haluts2023modelling}. These effects, we propose, could model a ubiquitous mechanism for contest resolution in real animal contests.

\clearpage

\section{Effective gravity model of swarming midges}\label{part2}

A special type of collective animal behavior is the formation of cohesive swarms \cite{couzin2009collective,ouellette2022physics}. These swarms are characterized by having the individual animals moving in an uncoordinated manner, unlike flocks, which are characterized by strong directional alignment, but nevertheless maintaining a cohesive structure that does not disperse. There have been various theoretical models proposed to describe this phenomenon, mainly based on velocity alignment that is mitigated by large noise \cite{cavagna2017dynamic,cavagna2018physics}. 

Here we describe a theoretical model that is based on adaptive long-range interactions~\cite{gorbonos2016long}. For swarming midges the effective interactions are attractive and mediated by acoustics due to the sound they emit while flying. Male midges form large swarms at dusk, thereby attracting females for mating. Within our model, midges form such cohesive swarms by accelerating towards each other in proportion to the intensity of the sound received. For pure acoustics the leading term (monopole) gives a functional form of the acceleration which is similar to the force of gravitational attraction, decaying as $1/r^2$. It was shown in~\cite{topaz2004swarming,topaz2006nonlocal} that in general power law decaying attraction and nonlinear diffusion are sufficient for the formation of local stable swarms from initially dispersed population profiles.

An additional, and crucial, component of this model is adaptivity. Any sensory mechanism in nature is subject to a modification due to adaptivity, which means that the sensory mechanism adapts itself to the background stimuli -- its sensitivity is higher when the background is lower and vice versa.
Exact adaptation means that the steady state output is independent of the steady-state level of input. This is part of a fold-change detection mechanism~\cite{shoval2010}, which is ubiquitous in nature, and involves a response whose entire shape, including amplitude and duration, depends only on fold change and not on the absolute levels of the input. The adaptivity is described by a scalar symmetry of the sensory mechanism whereby multiplication of the input fields or stimuli by a scalar does not change the reaction of the system.

Adding the two components together we reach the formulation of ``adaptive gravity'' which is described below. The adaptive-gravity model was shown to provide a good description of the observed mass and velocity profiles of laboratory midge swarms, which differ from those produced by regular gravity observed in star clusters~\cite{PhysRevResearch.2.013271}. Here we summarize some of the main theoretical results that describe the special features of the adaptive-gravity and appeared already in~\cite{gorbonos2020pair,PhysRevE.95.042405,PhysRevResearch.2.013271,gorbonos2016long}, focusing on the special properties that emerge due to these nonreciprocal long-range interactions. The parameters of the model are summarized in Table \ref{tabelParameters2}.

\begin{table}[h!]
  \begin{center}
  \begin{tabular}{m{0.7cm} m{2.5cm} m{9.5cm} m{2.7cm}}
    \toprule
    \textbf{Eq.} & \textbf{Parameter} & \textbf{Role} & \textbf{Dimensions} \\
    \midrule

    \eqref{FirstPotential} & $c_g$ & Strength of gravitational potential & $\mathrm{m l^3 t^{-2}}$ \\[1.5em]

    \eqref{AdapForce} & $R_{ad}$ & Length scale of adaptivity & $\mathrm{l}$  \\[1.5em]

    \eqref{FirstRs} & $R_{s}$ & The radius of the swarm - the mean distance of a particle from the center of mass & $\mathrm{l}$  \\[1.5em]

    \eqref{bgdDef} & $I_{bgd}$ & The leading order of the expansion of the "background noise" term & $\mathrm{m l t^{-2}}$  \\[1.5em]

    \eqref{BgdP} \eqref{step} & $\gamma, \gamma_1,\gamma_2$ & The background sound parameter & $-$ \\[0.5em]   
    \bottomrule

  \end{tabular}
  \end{center}
  \caption{\textbf{The parameters of the adaptive gravity model.} A list of all parameters used in the model's equations. The column '\textbf{Eq.}' indicates the equation in which each parameter is first used. The column '\textbf{Role}' provides a brief description of each parameter's role in the model. The column '\textbf{Dimensions}' gives the dimensions in terms of ($\mathrm{l - length}$, $\mathrm{m - mass}$, $\mathrm{t - time}$).}
  \label{tabelParameters2}
\end{table}

\subsection {Adaptive gravity} 
Let us consider N identical particles, whose positions are denoted by $\vec{x}_{i}$ ($i=1,..,N$), interacting by adaptive gravity. The contribution of a particle at $\vec{x}_{j}$ to the (regular) gravitational potential at $\vec{x}_{i}$ is given by
\begin{equation}
V^{ij}(\vec{x}_{i})=\begin{cases}
   \frac{c_{g}}{|\vec{x}_i-\vec{x}_j|}, & i\neq j\\
   0, & i=j.
\end{cases} \label{FirstPotential}
\end{equation}
where $c_{g}$ is a positive constant with dimensions of $mass\cdot length^3/time^2$.

The contribution of the particle at $\vec{x}_{j}$ to the gravitational force at the point $\vec{x}_i$ is given by the gradient of the potential
\begin{equation}
\vec{F}_{ij}^{g}(\vec{x}_{i})=-\nabla V^{ij}(\vec{x}_{i}),
\end{equation}
so that total gravitational (regular) linear force that acts on the particle at $\vec{x}_i$ is 
\begin{equation}
\vec{F}^{g} (\vec{x}_{i})=\sum_{j=1}^{n}\vec{F}_{ij}^{g}(\vec{x}_{i})=-\sum_{j=1}^{n}\nabla V^{ij}(\vec{x}_{i})
\end{equation}

In adaptive gravity, the regular gravitational force is ``renormalized'' by a factor that depends on the total background force (scalar sum in the denominator), resulting in a highly nonlinear form of the interactions between the particles
\begin{equation}
\vec{F}^{ad}(\vec{x}_{i})=-\frac{\sum_{j=1}^{n}\nabla V^{ij}(\vec{x}_{i})}{1+(c_{g}^{-1}R_{ad}^2)\sum_{j=1}^{n}|\nabla V^{ij}|} \label{AdapForce}
\end{equation}
where $R_{ad}$ is the length scale over which adaptivity occurs. This way to introduce adaptivity first appeared in the context of alignment interactions~\cite{motsch2011new,motsch2014heterophilious}. In the context of the midge swarm the effective interactions are mediated by acoustic (and also visual) signals, such that the background force represents the local background noise produced by the swarm. 

Since $|\nabla V^{ij}|\sim |\vec{x}_i-\vec{x}_j|^{-2}$, when the distances between pairs are large $|\vec{x}_i-\vec{x}_j|\gg \sqrt{N}R_{ad}$ the adaptivity does not play a role, and the interaction approaches regular gravity in this limit
\begin{equation}
\vec{F}^{ad}(\vec{x}_{i})\sim \vec{F}^{g}(\vec{x}_{i}).
\end{equation}
On the other hand, when 
$|\vec{x}_i-\vec{x}_j|<\sqrt{N}R_{ad}$ adaptivity is strong and then the following expression is a good approximation, which we term ``perfect adaptvity''
\begin{equation}
\vec{F}^{ad}(\vec{x}_{i})\sim \frac{c_g}{R_{ad}^2\,N_{tot}(\vec{x}_{i})}\,\sum_{j=1}^{n}\nabla V^{ij}(\vec{x}_{i}), \label{perfectAdap}
\end{equation}
where $N_{tot}(\vec{x}_{i})$ is a factor that represents the total background ``noise'' (or the total stimuli) at the point $\vec{x}_{i}$ as a result of all the interactions that involve the particle at this point
\begin{equation} \label{totalNoise}
N_{tot}(\vec{x}_{i})=\sum_{j=1}^{n}|\nabla V^{ij}(\vec{x}_{i})|.
\end{equation}
The appearance of this factor in the denominator of the expressions in Eqs.~(\ref{AdapForce}) and~(\ref{perfectAdap}) of the effective force, due to adaptivity, is responsible for the nonlinear and nonreciprocal nature of the interactions in this model. It is also a good example of the scalar symmetry in adaptation mechanisms~\cite{shoval2010}. We refer to $V^{ij}$ ($i,j= 1, .., N$) as our input fields and multiplication of the input fields by a scalar does not change the reaction of the system $\vec{F}^{ad}$.

For two particles at positions $\vec{x}_{1}$ and $\vec{x}_{2}$ this factor is reciprocal
\begin{equation}
N_{tot}(\vec{x}_{1})=N_{tot}(\vec{x}_{2})=\frac{c_g}{|\vec{x}_{1}-\vec{x}_{2}|},
\end{equation}
and therefore the interaction is reciprocal. Note that the effective force between two particles under perfect adaptivity becomes a constant, independent of the separation between them. Also, energy is conserved and the force from Eq.~(\ref{AdapForce}) can be integrated to give the effective potential between the pair of particles
\begin{equation}
U_{\mbox{\scriptsize pair}}(|\vec{x}_1-\vec{x}_2|)=\frac{c_g}{R_{\mbox{\scriptsize ad}}}\left[\arctan{(|\vec{x}_1-\vec{x}_2|/R_{\mbox{\scriptsize ad}})}-\pi/2\right].
\label{ueff}
\end{equation}

\begin{figure}[htb!]
    \centering
    \includegraphics[width =0.8\linewidth]{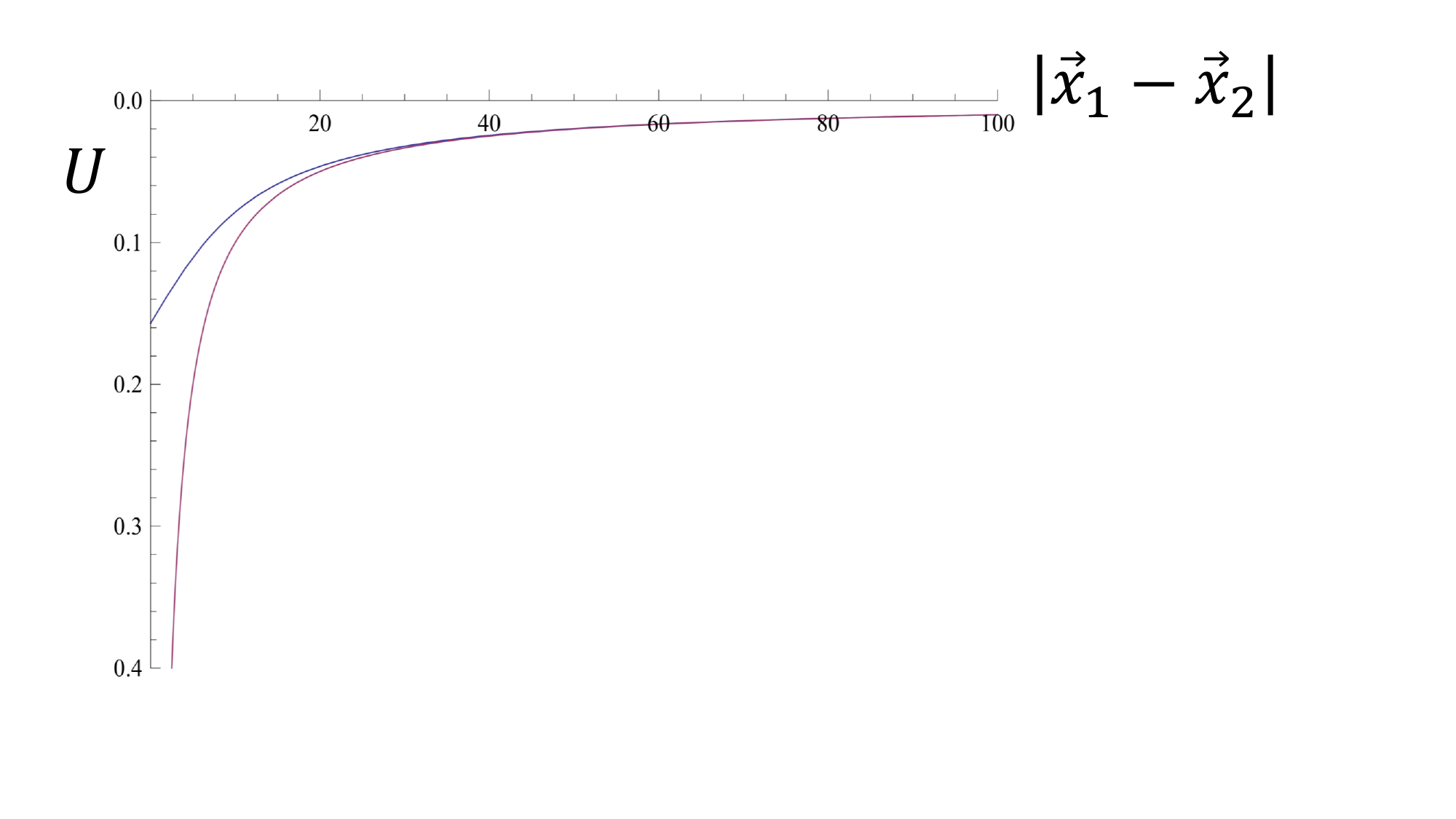}
    \caption{\textbf{The potential energy for two particles in adaptive gravity $U_{\mbox{\scriptsize pair}}(|\vec{x}_1-\vec{x}_2|)$ calculated according to Eq.~\ref{ueff}} using $R_{ad}/c_g=10$ (blue line). The red line is regular gravity potential for comparison with $c_g=1$.
    Adapted from~\cite{gorbonos2016long}.  
    }
    \label{AdaptivePairPotential}
\end{figure}

Unlike regular gravity, where the potential energy diverges when the two particles approach each other, the adaptivity prevents this and the potential in Eq.\ref{ueff} approaches a finite value with a ``cusp'' shape, so that it is linear in $|\vec{x}_1-\vec{x}_2|$. For $|\vec{x}_1-\vec{x}_2|\geq R_{\mbox{\scriptsize ad}}$ the effective potential approaches the long-range regular gravitational behavior of: $U_{\mbox{\scriptsize pair}}=c_{g}/|\vec{x}_1-\vec{x}_2|$.

In the case of $N \ge 3$ the interactions are in general nonreciprocal as we can see for three particles at $\vec{x}_{i}, \; (i=1,..,3)$, which have different normalization factors
\begin{eqnarray}
N_{tot}(\vec{x}_{1})&=&\frac{c_g}{|\vec{x}_{1}-\vec{x}_{2}|}+\frac{c_g}{|\vec{x}_{1}-\vec{x}_{3}|},\\
N_{tot}(\vec{x}_{2})&=&\frac{c_g}{|\vec{x}_{2}-\vec{x}_{1}|}+\frac{c_g}{|\vec{x}_{2}-\vec{x}_{3}|}\\
N_{tot}(\vec{x}_{3})&=&\frac{c_g}{|\vec{x}_{1}-\vec{x}_{3}|}+\frac{c_g}{|\vec{x}_{2}-\vec{x}_{3}|}
\end{eqnarray}

We therefore can not write an effective potential for this system. When Newton's third law is not satisfied, there is no conservation of momentum. In addition nonreciprocal interactions imply that the system is out of equilibrium and time reversibility is broken~\cite{Loos_2020}, and the energy cannot be conserved. Next we explore some of the notable consequences of the adaptive interactions on the dynamics and properties of the swarms.

\subsection{Enhanced stability of adaptive spherical-uniform swarms}

We now show that due to adaptivity a swarm of particles interacting through adaptive gravity does not suffer from the famous Jeans instability that is responsible for collapse of groups of particles under normal gravitational or other attractive power-law long range forces~\cite{Jeans,binney2011galactic}. Jeans instability means that any self-gravitating system that is dense enough is unstable to random fluctuations, which will result in the collapse of the system. In the case of biological matter, we expect that at short distances, the particles will repel each other, and the meaning of this instability could be a tearing off of the structure of the swarm. Nevertheless, adaptivity prevents it.  Adaptivity thus provides swarms with a natural mechanism for self-stabilization, which entails preserving their structure and preventing such instabilities. The argument that we present below is heuristic. A more rigorous derivation in terms of a critical wave number for a wavelike periodic perturbation appeared in~\cite{PhysRevE.95.042405} and reached the same expressions up to numerical constants. 

For the purpose of the argument let us consider a spherical uniform swarm whose radius is $R_s=\langle r \rangle$ (defined as the mean distance of a particle from the center of the mass of the swarm, which is $R_{s}=(3/4)R$ for a uniform density spherical swarm of radius $R$). We want to compare the response to a local fluctuation in the density for regular vs. adaptive gravity. In the case of a random fluctuation in the density, the random movement of the particles could stabilize the swarm if their random velocities are high enough. The typical time for stabilization is the typical time it takes to move across the swarm: $t_{esc}=R_s/\sigma_{v}$, where $\sigma_{v}=\sqrt{ \langle v^2 \rangle}$ is the root-mean-square velocity. This velocity can arise from thermal motion, or from the chaotic motion due to the attractive forces themselves (that depend on initial conditions), as well as due to the noisy active propulsion forces of the living (active) particles. The question is whether $\sigma_v$ is high enough so that the typical time to stabilize (or equivalently the time for the density fluctuation to escape) $t_{esc}$ is shorter than the typical time to collapse $t_{col}$, which is given by the typical time a particle falls to the center of mass due to the overall gravitational attraction: $t_{col}=2\,\pi/\sqrt{K}$, where $K$ is the effective spring constant for the linear restoring force of the form $\vec{F}=-K\vec{r}$ that acts towards the center of the swarm. This linear force appears for a uniform density spherical swarm, and is calculated in the appendix for both regular gravity $K^{g}$ (Eq.~(\ref{SpringRegular}) in the Appendix) and adaptive gravity $K^{ad}$ (Eq.~(\ref{SpringAdaptive}) in the Appendix). 

The criterion for instability is therefore 
$t_{esc}>t_{col}$. For regular gravity we get from it the following inequality
\begin{equation}
\rho>\frac{3\,\pi\,\langle v^2 \rangle}{c_g\,R_{s}^2}, \label{FirstRs}
\end{equation}
which gives the famous critical Jeans density for collapse in regular gravity \cite{Jeans}
\begin{equation}
\rho_{Jeans}=\frac{3\,\pi\,\langle v^2 \rangle}{c_g\,R_{s}^2},
\end{equation}
which means that for any density higher then this value the swarm will be unstable and collapse as a result of any arbitrarily small density fluctuation. 

In the case of adaptive gravity we get the following inequality:
\begin{equation}
\langle v^2 \rangle < \frac{c_g R_{s}}{16\,\pi^2\,R_{ad}^2},
\end{equation}
which shows that the collapse does not depend on the density, and there is a threshold on the value of the thermal velocity fluctuations below which the swarm will collapse. This means that any swarm that is ''hot'' enough is perfectly stable against density fluctuations. The adaptivity makes the mutual interactions weaker in the dense regions and as a result the dependence of stability on the density disappears, and this way it contributes to the stability of the swarm. 

The calculation given above for the stability condition of a uniform density swarm applies also for a real swarm (as well as a galaxy or a star-cluster) even though its density is not uniform \cite{PhysRevResearch.2.013271}.

It was found in~\cite{PhysRevE.95.042405} that for attractive forces with higher order power laws and adaptivity the stabilization is even stronger. Instead of a minimal density for collapse, there is a maximal density for collapse, and above it the swarm becomes stable, which therefore protects the swarms from collapse.

\subsection{Generalized virial relation for active systems}

We now develop virial equations that relate the kinetic energy to a generalized analogue of potential energy when the system is stationary, based only on the conservation of mass. The equations are of second order since they relate second-order moments. In their derivation we follow closely Chandrasekhar~\cite{Chandrasekhar} that derived them for the gravitational potential and we write them in a more general form for any force assuming only mass conservation. We derive here the scalar virial equation, while the more general tensorial version appeared in~\cite{gorbonos2016long}.

Let us start from the following version of Newton's second law governing the velocity $\vec{v}$ of an element of the swarm whose density is $\rho$:
\begin{equation}
\rho\,\frac{d \vec{v}}{dt}=-\nabla p+\rho\,\vec{f}, \label{fluid}
\end{equation}
where $p(\vec{r},t)$ is an isotropic pressure whose gradient creates a force and $\vec{f}$ stands for the rest of the forces per unit mass. We multiply the equation by the position vector $\vec{r}$ (scalar product) and integrate over the entire volume $V$. Then from the left hand side of Eq.~(\ref{fluid}) we obtain from mass conservation
\begin{equation}
\int \rho \vec{r}\cdot \frac{d \vec{v}}{dt} \,dV=\frac{d^2 I}{dt^2}-2\,K, \label{left}
\end{equation}
where $I$ is the scalar moment of inertia defined as
\begin{equation}
I\equiv\frac{1}{2}\int \rho \vec{r}^2\, dV,
\end{equation}
and 
\begin{equation}
K\equiv\frac{1}{2}\int \rho \vec{v}^2\, dV.
\end{equation}
The first term on the right hand side of Eq.~(\ref{fluid}) gives us:
\begin{equation}
\int \vec{r}\cdot \nabla p \, dV=\int \nabla \cdot (\vec{r} p)\, dV-3\int p\, dV.
\end{equation}
Then using Gauss's theorem for a volume $V$ and its boundary $\partial V$ we get a surface term for the pressure:
\begin{equation}
\int \vec{r}\cdot \nabla p\, dV=S-3\Pi, \label{right1}
\end{equation}
where the surface term is
\begin{equation}
S\equiv\oint_{\partial V}p\,\vec{r}\cdot d\vec{s},
\end{equation}
and we define the total pressure over the volume of the swarm to be:
\begin{equation}
\Pi \equiv \int p \,dV.
\end{equation}
The second term in the right hand side of Eq.~(\ref{fluid}) gives the analogue of potential energy of the swarm:
\begin{equation} \label{potential}
W\equiv\int \rho \vec{r}\cdot \vec{f} \,dV.
\end{equation}
Then combining Eqs. (\ref{fluid}),(\ref{left}) and (\ref{right1}) we obtain the following equation:
\begin{equation}
\frac{d^2 I}{dt^2}=2\,K-S+3\,\Pi+W. \label{AlmostVirial}
\end{equation}
Assuming that the system is stationary and defining the total analogue of kinetic energy $T$ as a sum of the kinematic motion and the internal thermal motions in the form of isotropic pressure:
\begin{equation}
T\equiv K+\frac{3}{2}\,\Pi,
\end{equation}
we arrive to a generalized virial theorem that relates (with a surface term) the kinetic energy with the ``potential energy'' W:
\begin{equation}
2\,T-S+W=0. \label{Virial}
\end{equation}
This relation is very general for any force $\vec{f}$ that acts on the particles without a requirement for any conservation laws except for the mass conservation. We can obtain average values for the analogues of kinetic and potential energies $T$ and $W$, by calculating the average of $\vec{v}^2$ and $\vec{r}\cdot \vec{f}$ (where $\vec{f}$ is taken as the acceleration towards the center of the swarm with the calculation being averaged per midge over the swarm volume and a long period of time).

Comparing this virial expression to data of real midge swarms in lab experiments the following mean values per midge were found~\cite{gorbonos2016long}:
\begin{eqnarray}
    \langle T \rangle=(3.42\pm0.08)\cdot 10^2\quad cm^2/s^2,\nonumber
\\
\langle-W/2\rangle=(2.80\pm0.08)\cdot 10^2\quad cm^2/s^2.
\end{eqnarray}
which were found to be approximately constant as a function of the swarm size $R_{s}$ (which increases with the number of midges in the swarm). Then according to Eq.~(\ref{Virial}), the difference between the two gives the mean surface pressure in the swarm:
\begin{equation}
S=(1.24\pm 0.14)\cdot 10^2 \quad cm^2/s^2.
\end{equation}
This pressure acts in the opposite direction to the kinetic energy in Eq.~\eqref{Virial}, indicating that the swarm is experiencing a stabilizing inwards effective pressure on its surface. The origin of this pressure could arise from interactions of the swarm midges with midges outside the swarm or the environment. Such external stabilizing pressures are commonly found in astrophysical stellar systems, such as globular clusters~\cite{Shapiro}.

For a spherical swarm of uniform density we can calculate the ``potential energy'' $W$, find its scaling with the swarm size $R_{s}$, and compare with the experimental data. We expect that near the center of the swarm the adaptivity will be the strongest and the ``perfect adaptivity'' regime will be the best approximation for the effective force between the particles. The ``potential energy'' (Eq.~\eqref{potential}) in a uniform spherical symmetric swarm is given by:
\begin{equation} \label{SphericalPotential}
W=\frac{27}{64\,R_{s}^3}\int_{0}^{\frac{4}{3}R_s}r^3\,f(r)\,dr.
\end{equation}

\begin{figure}[htb!]
    \centering
    \includegraphics[width =0.9\linewidth]{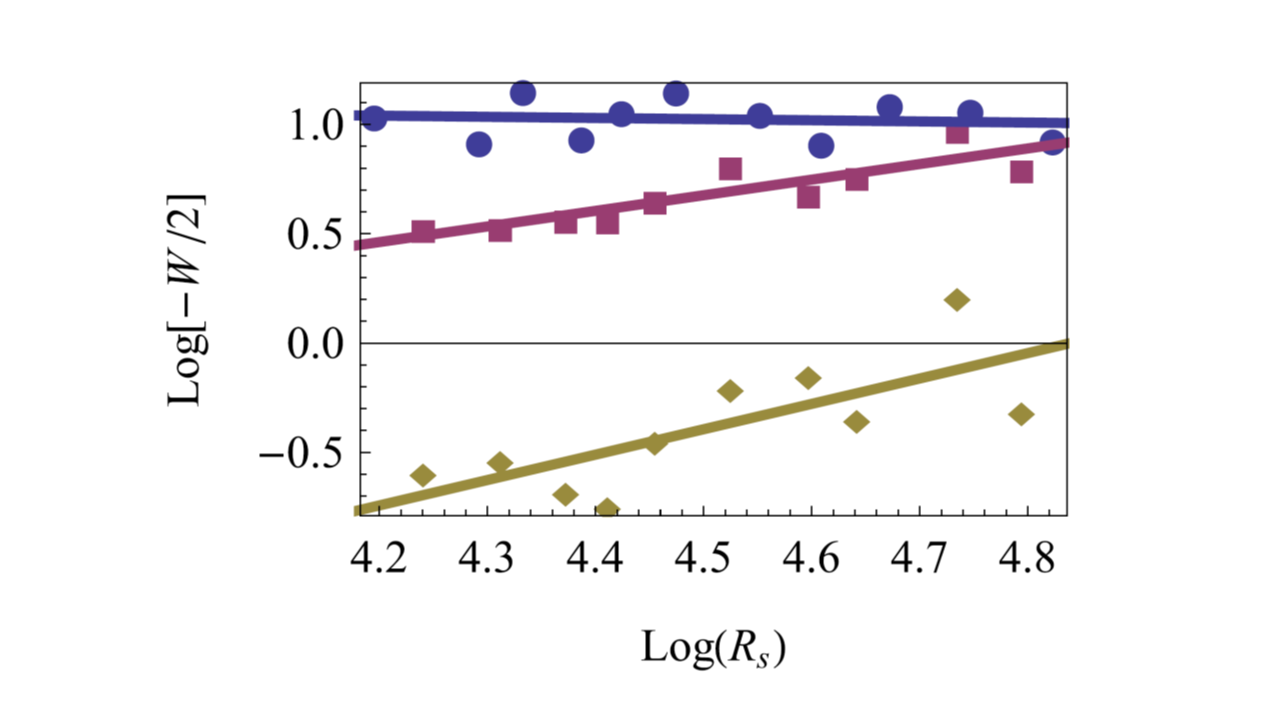}
    \caption{\textbf{The ``potential energy'' W in adaptive gravity} 
     Log-Log plot for the ``potential energy'' W in the center of the swarm with a linear fit, constrained for $r<R_s/2,R_s,\infty$ (yellow, purple and blue respectively). The power-law slopes are (yellow and purple respectively): $1.16\pm0.12, 0.71\pm0.05$.
    Adapted from \cite{gorbonos2016long}.}
    \label{AdaptivityInPotential}
\end{figure}

If it were a regular gravity then $f(r)=F^{g}(r)$ according to Eq.~$(\ref{New_grav})$ and
\begin{equation}
W^{g}=-\frac{9\,c_g}{20\,R_{s}} \propto -\frac{1}{R_{s}}.
\end{equation}
However, in adaptive gravity in the ``perfect adaptivity'' regime we have to divide $F^{g}(r)$ by $N_{tot}(r)$ according to the continuum limit of Eq.~(\ref{perfectAdap})
\begin{equation}
f(r)=\frac{c_g\,F^{g}(r)}{R_{ad}^2\,N_{tot}(r)}.
\end{equation}
Substituting into Eq.~(\ref{SphericalPotential}) and taking the expression for $N_{tot}(r)$ from Eq.~(\ref{total noise}) and $F^{g}(r)$ from Eq.~$(\ref{New_grav})$ we get:
\begin{equation}
    W^{ad}=-\frac{16\,c_g\,R_{s}}{3\,R_{ad}^2}\int_{0}^{1}\frac{x^5\,dx}{2x+(x^2-1)\ln{\left(\frac{1-x}{1+x}\right)}}\sim -0.4\frac{c_g}{R_{ad}^2}\,R_{s} \propto -R_{s},
\end{equation}
and this linear scaling with $R_{s}$ is indeed observed in the swarm data when we constrain the data closer to the center of the swarm (where the density is rather uniform and the effect of adaptivity is strongest), as we see from the slopes in Fig.~\ref{AdaptivityInPotential}.

Another way to look at Eq.~(\ref{AlmostVirial}) (when the system is stationary) is as a definition of the pressure:
\begin{equation}
\Pi=-\frac{1}{3}\left(2\,K+W\right)+\frac{S}{3},
\end{equation}
that was used to characterize the phases of midge swarms~\cite{sinhuber2017phase} and to write the equation of state for the swarms~\cite{sinhuber2021equation}.

\subsection{Spontaneous formation of bounded pairs in adaptive-gravity swarms}

\begin{figure}[htb!]
    \centering
    \includegraphics[width =\linewidth]{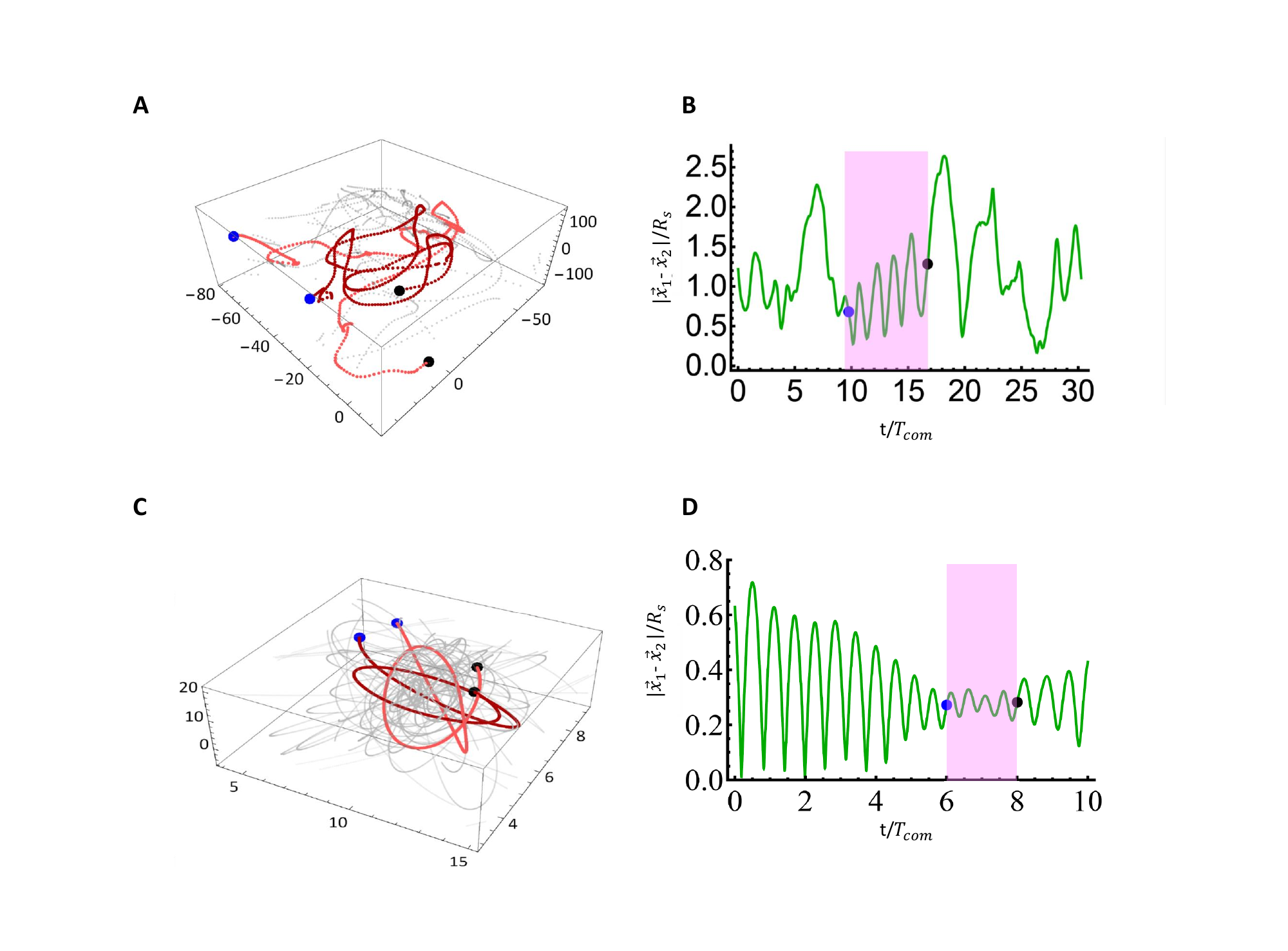}
    \caption{\textbf{Pair formation in laboratory observations of midge swarms
    } (\textbf{\textit{A}})   Trajectories of two midges in a laboratory swarm with 21 midges that exhibited
pairing (identified as being above a threshold value for the oscillation frequency, see~\cite{gorbonos2020pair,PhysRevLett.114.258103}). The midges were identified as belonging to a pair between
the blue and black points. Paired parts of the trajectories are colored in red, while unpaired
parts are in grey. Distances are in mm.
(\textbf{\textit{B}})  Distance between the members of the laboratory pair in (A) as a function of time. Distances are normalized by the swarm size $R_s$. The time is normalized by the typical orbit time around the center of mass $T_{com}$.
(\textbf{\textit{C}}) Pairing event in a simulation with 30 particles, $c_g=1$, $R_{ad}=15$ and $R_{s}=5.08$ (defined as a pair in a bound orbit). Distances are in simulation unit lengths. Here the segments between the blue and the black points are only a part of a longer pairing event. The red and the grey colors are as in (A).
(\textbf{\textit{D}})  Distance between the members of the pair in (C) (simulation) as a function of time. Distances are normalized by the swarm size $R_s$. The time is normalized by the typical orbit time around the center of mass $T_{com}$. Adapted from~\cite{gorbonos2020pair}.}
    \label{PairsLab}
\end{figure}

Besides the quantitative modifications as a result of adaptivity, which gives rise to modified global swarm behavior and scaling as shown above, there is a striking new local phenomenon of pair formation that can be qualitatively explained to arise from adaptivity. It has been observed that midges form transient pairs with synchronized relative motion while moving through the swarm~\cite{puckett2015}. An example from real data is shown in Fig.~\ref{PairsLab})(a,b) and from simulations in  
Fig.~\ref{PairsLab})(c,d). In the simulation we integrated the following equation (assuming a unit mass per particle)
\begin{equation}
\ddot{\vec{x}}_{i}=\vec{F}^{ad}(\vec{x}_{i})
\end{equation}
for all the particles in order to obtain an individual trajectory for each one of them, where $\vec{F}^{ad}(\vec{x}_{i})$ is given in Eq.~(\ref{AdapForce}). The definition of a pair in the data is according to the frequency of their relative movement being high enough, above a threshold value, and in the simulation the pair is defined by the energy ratio that forms a bound pair. We can see in the example of Fig.~\ref{PairsLab} that the bound pair in the simulation shows higher oscillation frequency in their mutual orbits and it is consistent with the definition of the pair in the laboratory data. It was also shown in~\cite{gorbonos2020pair} that this identification is statistically robust.

\begin{figure}[htb!]
    \centering
    \includegraphics[width =\linewidth]{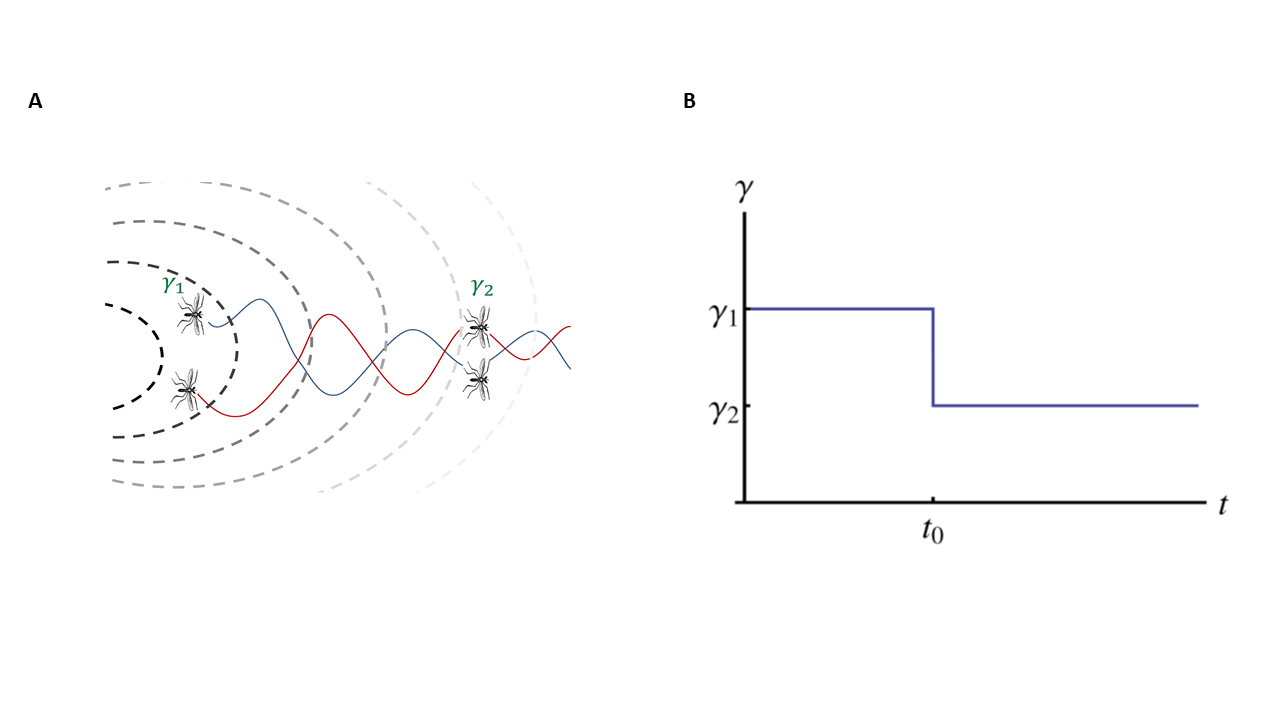}
    \caption{\textbf{The background sound parameter} (\textbf{\textit{A}}) Illustration of the proposed pair formation mechanism. When two interacting midges leave the dense region of the swarm (darker dashed lines), where the background sound parameter $\gamma$ is high, and move to a lower density region (such that $\gamma_1>\gamma_2$), the mutual pull between them becomes stronger, their orbit gets tighter, and they become bound. (\textbf{\textit{B}}) 
     The background sound parameter $\gamma$ is taken to be a step function. For $t\leq t_0$ the background sound level is higher than the background sound for $t> t_0$.
    Adapted from \cite{gorbonos2020pair}.}
    \label{BgdSound}
\end{figure}

Within the adaptive-gravity model these pairing events occur whenever two midges happen to move together from the center of the swarm  (where the total background noise is high) towards the swarm periphery (where the total background noise is low). As a result of adaptivity, during this movement, the attraction in the pair increases as the total noise decreases, thereby forming a bound state (see Fig.~\ref{BgdSound}A). The full analysis and comparison with experimental data appeared in~\cite{gorbonos2020pair}. Here we give a simple analytical sketch of the formation of a bound pair as a result of adaptivity. 

Let us consider the interaction of two midges, that happen by chance to come very close to each other, in the background of the rest. We assume that the separation between the two midges is small compared to their average distance to the rest of the midges in the swarm, and therefore the interactions with the rest of the swarm will be negligible except for a contribution to the total noise in Eq.~(\ref{AdapForce}), so that the effective force felt by one member of the pair (at $\vec{x}_1$) is

\begin{equation}
\vec{F}^{ad}(\vec{x}_{1})=-\frac{\nabla V^{12}(\vec{x}_{1})}{1+c_{g}^{-1}R_{ad}^2\,I_{background}+c_{g}^{-1}R_{ad}^2|\nabla V^{12}|}, \label{PairForce}
\end{equation}
where
\begin{equation} \label{bgdDef}
I_{background}\equiv\sum_{i=3}^{n}|\nabla V^{1i}|\sim I_{bgd}+\mathcal{O}.(|\vec{x}_{1}-\vec{x}_{i}|).
\end{equation}
The leading order of the expansion at infinity gives us a constant background contribution. Integrating this force we get the effective two-body potential (compare with Eq.~(\ref{ueff}))

\begin{equation}
U_{\mbox{\scriptsize pair}}(|\vec{x}_1-\vec{x}_2|)=\frac{c_g}{\gamma R_{\mbox{\scriptsize ad}}}\left[\arctan{(\gamma\,|\vec{x}_1-\vec{x}_2|/R_{\mbox{\scriptsize ad}})}-\pi/2\right],
\label{UeffBgd}
\end{equation}
where 
\begin{equation}
\gamma\equiv\sqrt{1+c_{g}^{-1}R_{ad}^2\,I_{bgd}} \label{BgdP}
\end{equation}
is the background sound parameter.
Hence we effectively have two-body motion under the influence of a mutual central force. In the case of two bodies, not only the conservation laws are valid but also the additivity of the effective force, so we can use all the conservation laws of a central force such as energy and angular momentum conservation. In particular, this two-body system can be reduced to an equivalent one-dimensional motion in the effective potential

\begin{equation} \label{gen_potential}
U_{\mbox{eff},12}(|\vec{x}_1-\vec{x}_2|)=\frac{\tilde{l}^2}{2\,|\vec{x}_1-\vec{x}_2|^2}+\frac{c_g}{\gamma\,R_{\mbox{\scriptsize ad}}}\left(\arctan\left(\frac{\gamma\,|\vec{x}_1-\vec{x}_2|}{R_{\mbox{\scriptsize ad}}}\right)-\frac{\pi}{2}\right),
\end{equation}
where $\tilde{l}$ is the angular momentum per unit mass (and we take the reduced mass of the pair).

When the background sound is reduced (as the pair move away from the swarm center), the sensitivity of each pair member increases, their effective mutual force becomes stronger and an unbound pair may become bound. Let us analyze it for simplicity when the background sound parameter decreases as a step function (Fig.~\ref{BgdSound}):
\begin{equation}
\gamma=\begin{cases}
      \gamma_1 & t  \leq t_0 \\
      \gamma_2 & t > t_0,
   \end{cases}
   \label{step}
\end{equation}
where $\gamma_1>\gamma_2$.

When the mutual attractive force is stronger, even when we start from an elliptical bound orbit, it becomes tighter and therefore the particles approach closer to each other. The total energy, which is conserved, is given by the following expression:
\begin{equation}
E_{i}=E_{k}(|\vec{x}_1-\vec{x}_2|)+\frac{\tilde{l}^2}{2\,|\vec{x}_1-\vec{x}_2|^2}+\frac{c_g}{\gamma_i\,R_{\mbox{\scriptsize ad}}}\left(\arctan\left(\frac{\gamma_i\,|\vec{x}_1-\vec{x}_2|}{R_{\mbox{\scriptsize ad}}}\right)-\frac{\pi}{2}\right),\end{equation}
where $E_{k}(|\vec{x}_1-\vec{x}_2|)$ is the kinetic energy of the radial movement. At $t=t_0$ let us denote the radial distance of the reduced mass by $r_0$ and the energy of the pair is changing due to the change in the background sound.
The energy difference is given by:
\begin{equation}
    \label{DifferenceEnergy}
    \Delta E =\Delta U_{\mbox{eff},12}=\frac{c_g}{\gamma_2\,R_{\mbox{\scriptsize ad}}}\left(\arctan\left(\frac{\gamma_2\,|\vec{x}_1-\vec{x}_2|}{R_{\mbox{\scriptsize ad}}}\right)-\frac{\pi}{2}\right)-\frac{c_g}{\gamma_1\,R_{\mbox{\scriptsize ad}}}\left(\arctan\left(\frac{\gamma_1\,|\vec{x}_1-\vec{x}_2|}{R_{\mbox{\scriptsize ad}}}\right)-\frac{\pi}{2}\right)
\end{equation}

When $\gamma_2<\gamma_1$ in this case $\Delta E <0$ since $U_{\mbox{eff},12}(r)$ which is given in Eq.~(\ref{gen_potential}) is monotonically increasing (becomes less negative) as a function of $\gamma$ for any positive value of $R_{ad}$
\begin{equation}
\partial_{\gamma} U_{\mbox{eff},12}=c_g\,\frac{2\,\gamma\,R_{ad}\,|\vec{x}_1-\vec{x}_2|+(R_{ad}^2+\gamma^2\,|\vec{x}_1-\vec{x}_2|^2)(\pi-2\,\arctan\left(\frac{\gamma\,|\vec{x}_1-\vec{x}_2|}{R_{ad}}\right))}{2\,\gamma^2\,R_{ad}\left(R_{ad}^2+\gamma^2\,|\vec{x}_1-\vec{x}_2|^2\right)}>0.
\end{equation}
The ``tightness'' of an orbit can be regarded as the absolute value of the ratio of the kinetic to potential energy $\left|E_k/ U_{\mbox{eff},12}\right|$. When it is smaller than one $\left|E_k/ U_{\mbox{eff},12}\right|<1$, the orbit is bound, and that is how a pair is defined, if the two midges move in bound orbits (see Fig.~\ref{PairsLab}). Since the effective potential energy is lowered as a result of the reduction of the background sound and the kinetic energy remains the same, this process can indeed produce bound orbits out of unbound ones. In addition, it can be shown~\cite{gorbonos2020pair} that the maximal distance between the members of the pairs is larger when the background sound parameter is higher. The proposed mechanism for pair formation is supported by experimental data~\cite{gorbonos2020pair}, where it is shown that the background sound gradient tends to be negative where the pairs form (upon leaving the swarm center) and tend to be positive where they dissociate (entering the swarm center).

This phenomenon of separated particles forming bounded pairs can be explained by the adaptive-gravity model, while in regular gravitational attraction it is extremely unlikely due to constraints of conservation of momentum and energy~\cite{bodenheimer2011principles}. Due to momentum conservation, two particles in regular gravity can form a bound pair only if a third particle removes the excess momentum. 

Note that the pairing mechanism that we describe is general to any many-body system with long-range adaptive interactions, and is not limited to the gravity-like functional form that we used here.

\subsection{Summary}

The fundamental ingredient of the adaptive gravity model is an ansatz for the effective force between particles. This force is a nonlinear function of the pairwise Newtonian gravitational interaction, inspired by the long-range acoustic interactions between midges in a swarm (Eq.~\eqref{AdapForce}):
\begin{equation*}
\vec{F}^{ad}(\vec{x}_{i})=-\frac{\sum_{j=1}^{n}\nabla V^{ij}(\vec{x}_{i})}{1+(c_{g}^{-1}R_{ad}^2)\sum_{j=1}^{n}|\nabla V^{ij}|}. 
\end{equation*}
This force is nonreciprocal for three particles or more. Among the various properties of the swarm in this model (that were compared to experimental results in~\cite{gorbonos2016long, PhysRevE.95.042405,PhysRevResearch.2.013271,gorbonos2020pair}) we chose to highlight here the following:
\begin{itemize}
    \item Stabilization of a swarm, based on the absence of gravitational Jeans instability, as a result of the adaptivity of the interactions.
    \item Generalized virial theorem which relates generalized kinetic energy $T$ with generalized potential energy $W$ (up to a surface term $S$) and is  a consequence of only mass conservation (Eq.~\eqref{Virial}):
    \begin{equation*}
2\,T-S+W=0. 
\end{equation*}
In addition we can predict the scaling relation between the effective potential energy and the swarm size (Fig. \ref{AdaptivityInPotential}).
    \item Spontaneous formation of pairs as a result of the adaptivity of the interactions, is another emergent feature of the long-range adaptive interactions.  
\end{itemize}

\section*{Discussion}

Nonreciprocal interactions between active particle has recently become a topic of growing interest. In this chapter, we have demonstrated that nonreciprocal forces, which arise from effective potentials that do not obey Newton's third law, and do not allow to formulate conservation of momentum and energy, arise naturally within the context of interactions between animals. We presented two very different examples for such effective interactions, in the form of (mostly) pairwise animal contests \cite{haluts2021spatiotemporal,haluts2023modelling}, and in the context of the many-body long-range interactions within a cohesive swarm \cite{gorbonos2016long, PhysRevE.95.042405,PhysRevResearch.2.013271,gorbonos2020pair}. These examples expand the types and scope of application of nonreciprocal interactions within the wider field of active matter.

The nonreciprocally interacting particles model of animal contests represents a departure from conventional approaches by explicitly addressing the spatio-temporal dynamics of contest behavior. Unlike traditional game-theoretic models \cite{payne1996escalation,mesterton1996wars,enquist1983evolution,enquist1990test,parker1981role,hammerstein1982asymmetric,payne1997animals,payne1998gradually,kokko2013dyadic}, which often overlook these dynamics, this model focuses on emergent spatio-temporal phenomena such as chase dynamics. By using effective interaction potentials distinct from traditional pairwise potentials, it captures the directional and nonreciprocal nature of contest interactions, highlighting the significance of asymmetric interactions in shaping agonistic behavior dynamics.

The effective gravity model of swarming midges introduces a unique perspective on collective behavior in animal groups. Moving beyond traditional velocity alignment-based models \cite{cavagna2017dynamic,cavagna2018physics}, this model leverages adaptive long-range interactions mediated by acoustics to explain swarm cohesion. The incorporation of adaptivity ensures sensitivity adjustment to background stimuli, leading to stable swarm formations and the spontaneous formation of pairs within the swarm. These findings underscore the importance of long-range interactions and adaptive mechanisms in maintaining cohesion within animal groups.

These models demonstrate that the study of active-particle systems with unconventional interactions enriches our understanding of animal behavior, while in turn the study of animal systems enriches and expands the scope of active-particle theory.

\appendix
\clearpage
\section*{Appendix}

\section{Explicit Calculation of the Linear Restoring Force of Adaptive Gravity}

\begin{figure}[tb]
\centering
\includegraphics[width=0.25\linewidth]{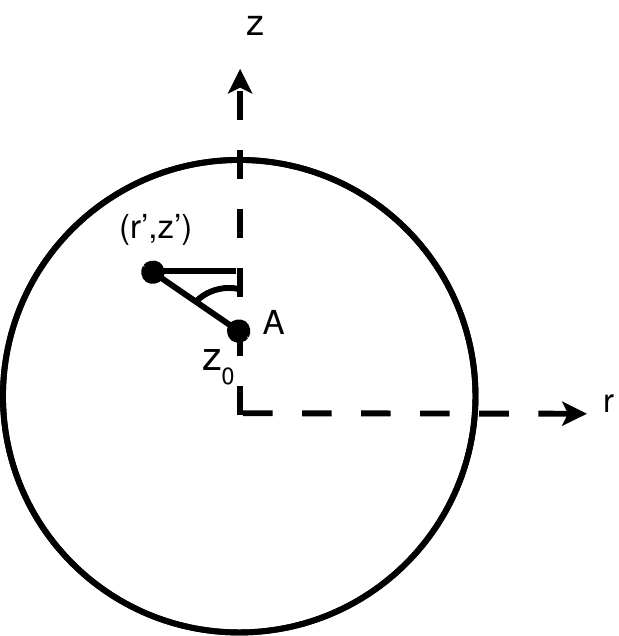}
\caption{\label{cylindricalcoordinates}  The cylindrical coordinates $(r,z)$ (and $\varphi$) that we use for the calculation of the effective gravitational field at a point A in a uniform-density spherical swarm.}
\end{figure}

Under the assumptions of a spherical and uniform swarm there is a linear restoring force towards the center of the swarm of the form $\vec{F}=-K\vec{r}$ in the leading order in the deviation from it (where $\vec{r}=0$ is the center of the swarm).
Let us calculate explicitly the effective spring constant of the linear adaptive-gravitational field at a point inside a spherical swarm with radius $R_s=(3/4)R$ (the mean distance from the center is $3/4$ of the radius $R$) and uniform density $\rho$, according to Eq.~(\ref{AdapForce}) where we take the continuum limit of the sums. We will use cylindrical coordinates $(r,z,\varphi)$ and calculate the field at  ($r=0$, $z=z_0$) without loss of generality (the point $A$ in Fig.~\ref{cylindricalcoordinates}). The symmetry of the problem implies that the field is along the $z$ axis. The contribution of a point at $(r',z')$ to the gravitational force at $(0,z_0)$ is
\begin{equation*}
    \frac{c_g}{r'^2+(z'-z_0)^2},
\end{equation*}
and the angle is
\begin{equation*}
\cos\varphi=\frac{z'-z_0}{\sqrt{r'^2+(z'-z_0)^2}}.
\end{equation*}
Hence the regular gravitational force at $z_0$ is
\begin{equation}
F^{g}(z_0)=2\,\pi\,\rho\,c_g\int_{-R}^{R}\!\!\!dz'\int_{0}^{\sqrt{R^2-z'^2}}\!\!\!\!\!r'dr'\,\frac{z'-z_0}{\left[r'^2+(z'-z_0)^2\right]^{\frac{3}{2}}} =-\frac{4\,\pi\,\rho\,c_g}{3}\,z_0,  \label{New_grav}
\end{equation}
and then the effective spring constant of this restoring force is 
\begin{equation}
K^{g}=\frac{4\,\pi\,\rho\,c_g}{3}. \label{SpringRegular}
\end{equation}
This result can be also obtained from Gauss's law (as it was derived in~\cite{gorbonos2016long}).
The ``total noise'' factor in the denominator (also in Eq.~\eqref{totalNoise}), which is given by the summation of the absolute values of the contributions to the point $(0,z_0)$, reads
\begin{equation}   
N_{\mbox{\scriptsize tot}}(z_0)=2\,\pi\,c_g\,\rho\,\int_{-R}^{R}dz'\int_{0}^{\sqrt{R^2-z'^2}}\frac{r'dr'}{r'^2+(z'-z_0)^2}
=\pi\,c_g\,\rho[2R-\frac{(R^{2}-z_0^{2})}{z_0}\ln\left(\frac{R-z_0}{R+z_0}\right)]. \label{total noise}
\end{equation}
To leading order in $z_0$, we have
\begin{equation}
N_{\mbox{\scriptsize tot}}(z_0)=4\pi\rho \,c_g\,R+\mathcal{O}(z_0^2), \label{Ntot}
\end{equation}
which after substitution into Eq.~(\ref{AdapForce}) gives to leading order in $z_0$ the following restoring force towards the center:
\begin{equation} \label{AdapRestor}
F^{ad}(z_0)=-\frac{4\,\pi\,\rho\,c_g}{3+16\,\pi\,\rho\,R_s\,R_{ad}^2}\,z_0.
\end{equation}

The number of particles $N$ in a uniform and spherical swarm is given by $N=4\,\pi \,\rho\,R_{s}^3/3$ and then we can write the denominator in Eq.~(\ref{AdapRestor}) in the following way
\[3+16\,\pi\,\rho\,R_s\,R_{ad}^2=3+16\,N\,R_{ad}^2/R_{s}^2.\]
Assuming the adaptivity range is approximately the size of the swarm $R_{ad} \sim R_s$ and $N \gg 1$. Then the first constant term in the denominator is negligible compared to the second. Then within this approximation we get the following expression for the effective spring constant:
\begin{equation}
K^{ad}=\frac{c_g}{4\,R_s\,R_{ad}^2}.
\label{SpringAdaptive}
\end{equation}

\renewcommand{\theequation}{A.\arabic{equation}}
\setcounter{equation}{0}
\printbibliography

@article{Jeans,
  title={The stability of a spherical nebula},
  author={J. H. Jeans},
  journal={Philosophical Transactions of the Royal Society of London, Series A},
  volume={199},
  number={},
  pages={1-53},
  year={1902},
  publisher={The Royal Society of London}
}

@book{binney2011galactic,
	Author = {Binney, James and Tremaine, Scott},
	Publisher = {Princeton university press},
	Title = {Galactic dynamics},
	Year = {2011}}

@article{mesterton1996wars,
  title={On wars of attrition without assessment},
  author={Mesterton-Gibbons, Michael and Marden, James H and Dugatkin, Lee Alan},
  journal={Journal of theoretical Biology},
  volume={181},
  number={1},
  pages={65--83},
  year={1996},
  publisher={Elsevier}
}

@article{payne1996escalation,
  title={Escalation and time costs in displays of endurance},
  author={Payne, Robert JH and Pagel, Mark},
  journal={Journal of Theoretical Biology},
  volume={183},
  number={2},
  pages={185--193},
  year={1996},
  publisher={Elsevier}
}

@article{arnott2009assessment,
  title={Assessment of fighting ability in animal contests},
  author={Arnott, Gareth and Elwood, Robert W},
  journal={Animal Behaviour},
  volume={77},
  number={5},
  pages={991--1004},
  year={2009},
  publisher={Elsevier}
}

@article{parker1974assessment,
  title={Assessment strategy and the evolution of fighting behaviour},
  author={Parker, Geoffrey A},
  journal={Journal of theoretical Biology},
  volume={47},
  number={1},
  pages={223--243},
  year={1974},
  publisher={Elsevier}
}

@article{enquist1983evolution,
  title={Evolution of fighting behaviour: decision rules and assessment of relative strength},
  author={Enquist, Magnus and Leimar, Olof},
  journal={Journal of theoretical Biology},
  volume={102},
  number={3},
  pages={387--410},
  year={1983},
  publisher={Elsevier}
}

@article{enquist1990test,
  title={A test of the sequential assessment game: fighting in the cichlid fish Nannacara anomala},
  author={Enquist, Magnus and Leimar, Olof and Ljungberg, Tomas and Mallner, Ylva and Segerdahl, Nils},
  journal={Animal Behaviour},
  volume={40},
  number={1},
  pages={1--14},
  year={1990},
  publisher={Elsevier}
}

@article{parker1981role,
  title={Role assessment, reserve strategy, and acquisition of information in asymmetric animal conflicts},
  author={Parker, Geoffrey A and Rubenstein, Daniel I},
  journal={Animal behaviour},
  volume={29},
  number={1},
  pages={221--240},
  year={1981},
  publisher={Elsevier}
}

@article{arnott2008information,
  title={Information gathering and decision making about resource value in animal contests},
  author={Arnott, Gareth and Elwood, Robert W},
  journal={Animal Behaviour},
  volume={76},
  number={3},
  pages={529--542},
  year={2008},
  publisher={Elsevier}
}

@article{taylor2003mismeasure,
  title={The mismeasure of animal contests},
  author={Taylor, PW and Elwood, Robert W},
  journal={Animal Behaviour},
  volume={65},
  number={6},
  pages={1195--1202},
  year={2003},
  publisher={Elsevier}
}

@article{romanczuk2012active,
  title={Active brownian particles},
  author={Romanczuk, Pawel and B{\"a}r, Markus and Ebeling, Werner and Lindner, Benjamin and Schimansky-Geier, Lutz},
  journal={The European Physical Journal Special Topics},
  volume={202},
  number={1},
  pages={1--162},
  year={2012},
  publisher={Springer}
}

@article{volpe2014simulation,
  title={Simulation of the active Brownian motion of a microswimmer},
  author={Volpe, Giorgio and Gigan, Sylvain and Volpe, Giovanni},
  journal={American Journal of Physics},
  volume={82},
  number={7},
  pages={659--664},
  year={2014},
  publisher={AAPT}
}

@article{vollrath1980male,
  title={Male body size and fitness in the web-building spider Nephila clavipes},
  author={Vollrath, Fritz},
  journal={Zeitsch-rift f{\"u}r Tierpsychologie},
  volume={53},
  number={1},
  pages={61--78},
  year={1980},
  publisher={Wiley Online Library}
}

@article{jordan2014reproductive,
  title={Reproductive foragers: male spiders choose mates by selecting among competitive environments},
  author={Jordan, Lyndon Alexander and Kokko, Hanna and Kasumovic, Michael},
  journal={The American Naturalist},
  volume={183},
  number={5},
  pages={638--649},
  year={2014},
  publisher={University of Chicago Press Chicago, IL}
}

@article{payne1998gradually,
  title={Gradually escalating fights and displays: the cumulative assessment model},
  author={Payne, Robert JH},
  journal={Animal Behaviour},
  volume={56},
  number={3},
  pages={651--662},
  year={1998},
  publisher={Elsevier}
}

@article{hammerstein1982asymmetric,
  title={The asymmetric war of attrition},
  author={Hammerstein, Peter and Parker, Geoffrey A},
  journal={Journal of Theoretical Biology},
  volume={96},
  number={4},
  pages={647--682},
  year={1982},
  publisher={Elsevier}
}

@article{haluts2021spatiotemporal,
  title={Spatiotemporal dynamics of animal contests arise from effective forces between contestants},
  author={Haluts, Amir and Reyes, Sylvia F Garza and Gorbonos, Dan and Etheredge, Robert Ian and Jordan, Alex and Gov, Nir S},
  journal={Proceedings of the National Academy of Sciences},
  volume={118},
  number={49},
  year={2021},
  publisher={National Acad Sciences}
}

@article{pinto2019all,
  title={All by myself? Meta-analysis of animal contests shows stronger support for self than for mutual assessment models},
  author={Pinto, Nelson S and Palaoro, Alexandre V and Peixoto, Paulo EC},
  journal={Biological Reviews},
  volume={94},
  number={4},
  pages={1430--1442},
  year={2019},
  publisher={Wiley Online Library}
}

@article{payne1997animals,
  title={Why do animals repeat displays?},
  author={Payne, Robert JH and Pagel, Mark},
  journal={Animal Behaviour},
  volume={54},
  number={1},
  pages={109--119},
  year={1997},
  publisher={Elsevier}
}

@article{gorbonos2016long,
  title={Long-range acoustic interactions in insect swarms: an adaptive gravity model},
  author={Gorbonos, Dan and Ianconescu, Reuven and Puckett, James G and Ni, Rui and Ouellette, Nicholas T and Gov, Nir S},
  journal={New Journal of Physics},
  volume={18},
  number={7},
  pages={073042},
  year={2016},
  publisher={IOP Publishing}
}

@article{hanggi1990reaction,
  title={Reaction-rate theory: fifty years after Kramers},
  author={H{\"a}nggi, Peter and Talkner, Peter and Borkovec, Michal},
  journal={Reviews of modern physics},
  volume={62},
  number={2},
  pages={251},
  year={1990},
  publisher={APS}
}

@article{smith1973logic,
  title={The logic of animal conflict},
  author={Smith, JMPGR and Price, George R},
  journal={Nature},
  volume={246},
  number={5427},
  pages={15--18},
  year={1973},
  publisher={Nature Publishing Group}
}

@article{smith1974theory,
  title={The theory of games and the evolution of animal conflicts},
  author={Smith, J Maynard},
  journal={Journal of theoretical biology},
  volume={47},
  number={1},
  pages={209--221},
  year={1974},
  publisher={Elsevier}
}

@article{smith1976logic,
  title={The logic of asymmetric contests},
  author={Smith, John Maynard and Parker, Geoffrey A},
  journal={Animal behaviour},
  volume={24},
  number={1},
  pages={159--175},
  year={1976},
  publisher={Elsevier}
}

@article{parker1979sexual,
  title={Sexual selection and sexual conflict},
  author={Parker, Geoffrey Alan and others},
  journal={Sexual selection and reproductive competition in insects},
  volume={123},
  pages={166},
  year={1979}
}

@incollection{kokko2013dyadic,
  title={Dyadic contests: modelling flights between two individuals},
  author={Kokko, Hanna and others},
  booktitle={Animal contests},
  year={2013},
  publisher={Cambridge University Press}
}

@article{nathan2022big,
  title={Big-data approaches lead to an increased understanding of the ecology of animal movement},
  author={Nathan, Ran and Monk, Christopher T and Arlinghaus, Robert and Adam, Timo and Al{\'o}s, Josep and Assaf, Michael and Baktoft, Henrik and Beardsworth, Christine E and Bertram, Michael G and Bijleveld, Allert I and others},
  journal={Science},
  volume={375},
  number={6582},
  pages={eabg1780},
  year={2022},
  publisher={American Association for the Advancement of Science}
}

@article{emlen1977ecology,
  title={Ecology, sexual selection, and the evolution of mating systems},
  author={Emlen, Stephen T and Oring, Lewis W},
  journal={Science},
  volume={197},
  number={4300},
  pages={215--223},
  year={1977},
  publisher={American Association for the Advancement of Science}
}

@incollection{fuxjager2017animals,
  title={Why animals fight: Uncovering the function and mechanisms of territorial aggression.},
  author={Fuxjager, Matthew J and Zhao, Xin and Rieger, Nathaniel S and Marler, Catherine A},
  booktitle={APA handbook of comparative psychology: Basic concepts, methods, neural substrate, and behavior, Vol. 1},
  pages={853--875},
  year={2017},
  publisher={American Psychological Association}
}

@article{rittschof2010male,
  title={Male density affects large-male advantage in the golden silk spider, Nephila clavipes},
  author={Rittschof, Clare C},
  journal={Behavioral Ecology},
  volume={21},
  number={5},
  pages={979--985},
  year={2010},
  publisher={Oxford University Press}
}

@article{schneider2000sperm,
  title={Sperm competition and small size advantage for males of the golden orb-web spider Nephila edulis},
  author={Schneider, JM and Herberstein, ME and De Crespigny, F Champion and Ramamurthy, S and Elgar, MA},
  journal={Journal of Evolutionary Biology},
  volume={13},
  number={6},
  pages={939--946},
  year={2000},
  publisher={Wiley Online Library}
}

@article{bellomo2022life,
  title={What is life? Active particles tools towards behavioral dynamics in social-biology and economics},
  author={Bellomo, N and Esfahanian, M and Secchini, V and Terna, P},
  journal={Physics of Life Reviews},
  year={2022},
  publisher={Elsevier}
}

@article{zottl2023modeling,
  title={Modeling active colloids: From active brownian particles to hydrodynamic and chemical fields},
  author={Z{\"o}ttl, Andreas and Stark, Holger},
  journal={Annual Review of Condensed Matter Physics},
  volume={14},
  pages={109--127},
  year={2023},
  publisher={Annual Reviews}
}

@article{dell2014automated,
  title={Automated image-based tracking and its application in ecology},
  author={Dell, Anthony I and Bender, John A and Branson, Kristin and Couzin, Iain D and de Polavieja, Gonzalo G and Noldus, Lucas PJJ and P{\'e}rez-Escudero, Alfonso and Perona, Pietro and Straw, Andrew D and Wikelski, Martin and others},
  journal={Trends in ecology \& evolution},
  volume={29},
  number={7},
  pages={417--428},
  year={2014},
  publisher={Elsevier}
}

@article{shklarsh2011smart,
  title={Smart swarms of bacteria-inspired agents with performance adaptable interactions},
  author={Shklarsh, Adi and Ariel, Gil and Schneidman, Elad and Ben-Jacob, Eshel},
  journal={PLoS computational biology},
  volume={7},
  number={9},
  pages={e1002177},
  year={2011},
  publisher={Public Library of Science San Francisco, USA}
}

@article{ben1992adaptive,
  title={Adaptive self-organization during growth of bacterial colonies},
  author={Ben-Jacob, Eshel and Shmueli, Haim and Shochet, Ofer and Tenenbaum, Adam},
  journal={Physica A: Statistical Mechanics and its Applications},
  volume={187},
  number={3-4},
  pages={378--424},
  year={1992},
  publisher={Elsevier}
}

@article{vicsek2012collective,
  title={Collective motion},
  author={Vicsek, Tam{\'a}s and Zafeiris, Anna},
  journal={Physics reports},
  volume={517},
  number={3-4},
  pages={71--140},
  year={2012},
  publisher={Elsevier}
}

@article{barberis2019phase,
  title={Phase separation and emergence of collective motion in a one-dimensional system of active particles},
  author={Barberis, Lucas and Peruani, Fernando},
  journal={The Journal of chemical physics},
  volume={150},
  number={14},
  year={2019},
  publisher={AIP Publishing}
}

@article{haskovec2013flocking,
  title={Flocking dynamics and mean-field limit in the Cucker--Smale-type model with topological interactions},
  author={Haskovec, Jan},
  journal={Physica D: Nonlinear Phenomena},
  volume={261},
  pages={42--51},
  year={2013},
  publisher={Elsevier}
}

@inproceedings{briffa2013introduction,
  title={Introduction to animal contests},
  author={Briffa, Mark and Hardy, Ian CW},
  booktitle={Animal contests},
  pages={1--4},
  year={2013},
  publisher={Cambridge University Press Cambridge, UK}
}

@article{chapin2019further,
  title={Further mismeasures of animal contests: a new framework for assessment strategies},
  author={Chapin, Kenneth James and Peixoto, Paulo Enrique Cardoso and Briffa, Mark},
  journal={Behavioral Ecology},
  volume={30},
  number={5},
  pages={1177--1185},
  year={2019},
  publisher={Oxford University Press UK}
}

@article{adler2018fold,
  title={Fold-change detection in biological systems},
  author={Adler, Miri and Alon, Uri},
  journal={Current Opinion in Systems Biology},
  volume={8},
  pages={81--89},
  year={2018},
  publisher={Elsevier}
}

@article{haluts2023modelling,
  title={Modelling animal contests based on spatio-temporal dynamics},
  author={Haluts, Amir and Jordan, Alex and Gov, Nir S},
  journal={Journal of the Royal Society Interface},
  volume={20},
  number={202},
  pages={20220866},
  year={2023},
  publisher={The Royal Society}
}

@article{paton1986communication,
  title={Communication by agonistic displays},
  author={Paton, D and Caryl, PG},
  journal={Behaviour},
  volume={98},
  number={1-4},
  pages={213--239},
  year={1986},
  publisher={Brill}
}

@article{martin2007review,
  title={A review of shark agonistic displays: comparison of display features and implications for shark--human interactions},
  author={Martin, R Aidan},
  journal={Marine and Freshwater Behaviour and Physiology},
  volume={40},
  number={1},
  pages={3--34},
  year={2007},
  publisher={Taylor \& Francis}
}

@article{popp1987risk,
  title={Risk and effectiveness in the use of agonistic displays by American goldfinches},
  author={Popp, James W},
  journal={Behaviour},
  volume={103},
  number={1-3},
  pages={141--156},
  year={1987},
  publisher={Brill}
}

@article{Loos_2020,
year = {2020},
month = {12},
publisher = {IOP Publishing},
volume = {22},
number = {12},
pages = {123051},
author = {Sarah A M Loos and Sabine H L Klapp},
title = {Irreversibility, heat and information flows induced by non-reciprocal interactions},
journal = {New Journal of Physics}
}

@article{motsch2011new,
  title={A new model for self-organized dynamics and its flocking behavior},
  author={Motsch, Sebastien and Tadmor, Eitan},
  journal={Journal of Statistical Physics},
  volume={144},
  pages={923--947},
  year={2011},
  publisher={Springer}
}

@article{motsch2014heterophilious,
  title={Heterophilious dynamics enhances consensus},
  author={Motsch, Sebastien and Tadmor, Eitan},
  journal={SIAM review},
  volume={56},
  number={4},
  pages={577--621},
  year={2014},
  publisher={SIAM}
}

@article{PhysRevE.95.042405,
  title = {Stable swarming using adaptive long-range interactions},
  author = {Gorbonos, Dan and Gov, Nir S.},
  journal = {Phys. Rev. E},
  volume = {95},
  issue = {4},
  pages = {042405},
  numpages = {10},
  year = {2017},
  month = {4},
  publisher = {American Physical Society},
}

@article{ouellette2022physics,
  title={A physics perspective on collective animal behavior},
  author={Ouellette, Nicholas T},
  journal={Physical Biology},
  volume={19},
  number={2},
  pages={021004},
  year={2022},
  publisher={IOP Publishing}
}

@article{cavagna2018physics,
  title={The physics of flocking: Correlation as a compass from experiments to theory},
  author={Cavagna, Andrea and Giardina, Irene and Grigera, Tom{\'a}s S},
  journal={Physics Reports},
  volume={728},
  pages={1--62},
  year={2018},
  publisher={Elsevier}
}

@article{cavagna2017dynamic,
  title={Dynamic scaling in natural swarms},
  author={Cavagna, Andrea and Conti, Daniele and Creato, Chiara and Del Castello, Lorenzo and Giardina, Irene and Grigera, Tomas S and Melillo, Stefania and Parisi, Leonardo and Viale, Massimiliano},
  journal={Nature Physics},
  volume={13},
  number={9},
  pages={914--918},
  year={2017},
  publisher={Nature Publishing Group UK London}
}

@article{couzin2009collective,
  title={Collective cognition in animal groups},
  author={Couzin, Iain D},
  journal={Trends in cognitive sciences},
  volume={13},
  number={1},
  pages={36--43},
  year={2009},
  publisher={Elsevier}
}

@article{PhysRevResearch.2.013271,
  title = {Similarities between insect swarms and isothermal globular clusters},
  author = {Gorbonos, Dan and van der Vaart, Kasper and Sinhuber, Michael and Puckett, James G. and Reynolds, Andrew M. and Ouellette, Nicholas T. and Gov, Nir S.},
  journal = {Phys. Rev. Res.},
  volume = {2},
  issue = {1},
  pages = {013271},
  numpages = {11},
  year = {2020},
  month = {3},
  publisher = {American Physical Society},
}

@article{PhysRevLett.114.258103,
  title = {Time-Frequency Analysis Reveals Pairwise Interactions in Insect Swarms},
  author = {Puckett, James G. and Ni, Rui and Ouellette, Nicholas T.},
  journal = {Phys. Rev. Lett.},
  volume = {114},
  issue = {25},
  pages = {258103},
  numpages = {5},
  year = {2015},
  month = {Jun},
  publisher = {American Physical Society},
  doi = {10.1103/PhysRevLett.114.258103},
  url = {https://link.aps.org/doi/10.1103/PhysRevLett.114.258103}
}

@article{shoval2010,
	Author = {O. Shoval and L. Goentoro and Y. Hart and A. Mayo and E. Sontag and U. Alon},
	Journal = {Proc. Natl. Acad. Sci. USA},
	Pages = {15995-16000},
	Title = {Fold-change detection and scalar symmetry of sensory input fields},
	Volume = {107},
	Year = {2010}}

@book{bodenheimer2011principles,
  title={Principles of Star Formation},
  author={Bodenheimer, P.},
  isbn={9783642150623},
  lccn={2011933131},
  series={Astronomy and Astrophysics Library},
  url={https://books.google.co.il/books?id=ZlfANAEACAAJ},
  year={2011},
  publisher={Springer Berlin Heidelberg}
}

@article{gorbonos2020pair,
  title={Pair formation in insect swarms driven by adaptive long-range interactions},
  author={Gorbonos, Dan and Puckett, James G and van der Vaart, Kasper and Sinhuber, Michael and Ouellette, Nicholas T and Gov, Nir S},
  journal={Journal of the Royal Society Interface},
  volume={17},
  number={171},
  pages={20200367},
  year={2020},
  publisher={The Royal Society}
}

@book{Chandrasekhar,
      Author = {Chandrasekhar, Subrahmanyan},
      Title         = {Ellipsoidal figures of equilibrium},
      Publisher     = {Yale Univ. Press},
      Address       = {New Haven, CT},
      Series        = {Silliman Foundati Lect.},
      Year          = {1969}}

@article{sinhuber2021equation,
  title={An equation of state for insect swarms},
  author={Sinhuber, Michael and van der Vaart, Kasper and Feng, Yenchia and Reynolds, Andrew M and Ouellette, Nicholas T},
  journal={Scientific Reports},
  volume={11},
  number={1},
  pages={3773},
  year={2021},
  publisher={Nature Publishing Group UK London}
}

@article{sinhuber2017phase,
  title={Phase coexistence in insect swarms},
  author={Sinhuber, Michael and Ouellette, Nicholas T},
  journal={Physical review letters},
  volume={119},
  number={17},
  pages={178003},
  year={2017},
  publisher={APS}
}

@article{Shapiro,
      author         = "Shapiro, Paul R. and Iliev, Ilian T. and Martel, Hugo and
                        Ahn, Kyungjin and Alvarez, Marcelo A.",
      title          = "{The Equilibrium structure of CDM halos}",
      year           = "2004",
      eprint         = "astro-ph/0409173",
      archivePrefix  = "arXiv",
      primaryClass   = "astro-ph",
      SLACcitation   = "%%CITATION = ASTRO-PH/0409173;%%"
}

@article{geiseler2016kramers,
  title={Kramers escape of a self-propelled particle},
  author={Geiseler, Alexander and H{\"a}nggi, Peter and Schmid, Gerhard},
  journal={The European Physical Journal B},
  volume={89},
  pages={1--7},
  year={2016},
  publisher={Springer}
}

@article{puckett2015,
	Author = {J. G. Puckett and R. Ni and N. T. Ouellette},
	Journal = {Phys. Rev. Lett.},
	Pages = {258103},
	Title = {Time-frequency analysis reveals pairwise interactions in insect swarms},
	Volume = {114},
	Year = {2015}}

@article{fruchart2021non,
  title={Non-reciprocal phase transitions},
  author={Fruchart, Michel and Hanai, Ryo and Littlewood, Peter B and Vitelli, Vincenzo},
  journal={Nature},
  volume={592},
  number={7854},
  pages={363--369},
  year={2021},
  publisher={Nature Publishing Group UK London}
}

@article{durve2018active,
  title={Active particle condensation by non-reciprocal and time-delayed interactions},
  author={Durve, Mihir and Saha, Arnab and Sayeed, Ahmed},
  journal={The European Physical Journal E},
  volume={41},
  pages={1--9},
  year={2018},
  publisher={Springer}
}

@article{kreienkamp2022clustering,
  title={Clustering and flocking of repulsive chiral active particles with non-reciprocal couplings},
  author={Kreienkamp, Kim L and Klapp, Sabine HL},
  journal={New Journal of Physics},
  volume={24},
  number={12},
  pages={123009},
  year={2022},
  publisher={IOP Publishing}
}

@article{knevzevic2022collective,
  title={Collective motion of active particles exhibiting non-reciprocal orientational interactions},
  author={Kne{\v{z}}evi{\'c}, Milo{\v{s}} and Welker, Till and Stark, Holger},
  journal={Scientific Reports},
  volume={12},
  number={1},
  pages={19437},
  year={2022},
  publisher={Nature Publishing Group UK London}
}

@article{dinelli2023non,
  title={Non-reciprocity across scales in active mixtures},
  author={Dinelli, Alberto and O’Byrne, J{\'e}r{\'e}my and Curatolo, Agnese and Zhao, Yongfeng and Sollich, Peter and Tailleur, Julien},
  journal={Nature Communications},
  volume={14},
  number={1},
  pages={7035},
  year={2023},
  publisher={Nature Publishing Group UK London}
}

@article{saha2020scalar,
  title={Scalar active mixtures: The nonreciprocal Cahn-Hilliard model},
  author={Saha, Suropriya and Agudo-Canalejo, Jaime and Golestanian, Ramin},
  journal={Physical Review X},
  volume={10},
  number={4},
  pages={041009},
  year={2020},
  publisher={APS}
}

@article{you2020nonreciprocity,
  title={Nonreciprocity as a generic route to traveling states},
  author={You, Zhihong and Baskaran, Aparna and Marchetti, M Cristina},
  journal={Proceedings of the National Academy of Sciences},
  volume={117},
  number={33},
  pages={19767--19772},
  year={2020},
  publisher={National Acad Sciences}
}

@article{corless1996lambert,
  title={On the Lambert W function},
  author={Corless, Robert M and Gonnet, Gaston H and Hare, David EG and Jeffrey, David J and Knuth, Donald E},
  journal={Advances in Computational mathematics},
  volume={5},
  pages={329--359},
  year={1996},
  publisher={Springer}
}

@article{topaz2004swarming,
  title={Swarming patterns in a two-dimensional kinematic model for biological groups},
  author={Topaz, Chad M and Bertozzi, Andrea L},
  journal={SIAM Journal on Applied Mathematics},
  volume={65},
  number={1},
  pages={152--174},
  year={2004},
  publisher={SIAM}
}

@article{topaz2006nonlocal,
  title={A nonlocal continuum model for biological aggregation},
  author={Topaz, Chad M and Bertozzi, Andrea L and Lewis, Mark A},
  journal={Bulletin of mathematical biology},
  volume={68},
  pages={1601--1623},
  year={2006},
  publisher={Springer}
}

@article{giardina2008collective,
  title={Collective behavior in animal groups: theoretical models and empirical studies},
  author={Giardina, Irene},
  journal={HFSP journal},
  volume={2},
  number={4},
  pages={205--219},
  year={2008},
  publisher={Taylor \& Francis}
}

\end{document}